\documentclass[12pt]{article}

\renewcommand{\thefootnote}{\fnsymbol{footnote}}

\usepackage{amsbsy,amssymb,latexsym,amsfonts,amsmath}
\usepackage{mathrsfs}
\usepackage{graphicx}
\usepackage{bm}
\usepackage{here}
\usepackage{comment}
\usepackage{amsmath}
\usepackage{cases}
\usepackage {empheq}
\usepackage{color}
\makeatletter
\def\EqNumText{\refstepcounter{equation}\cdots\tagform@\theequation}%
\makeatother

\newcommand{\bel}[1]{\begin{equation}\label{#1}}                     
\newcommand{\bal}[1]{\begin{eqnarray}\label{#1}}                     
\newcommand{\be}{\begin{equation}}
\newcommand{\ee}{\end{equation}}

\renewcommand\labelenumi{(\theenumi)}

\begin{document}
%
%
\begin{titlepage}
\begin{flushright}
\normalsize
~~~~
NITEP 63\\
OCU-PHYS 517\\
25 March, 2020 \\
\end{flushright}

\vspace{15pt}

\begin{center}
{\LARGE Stability, enhanced gauge symmetry and suppressed cosmological constant in 9D heterotic interpolating models} \\
\end{center}

\vspace{23pt}

\begin{center}
{ H. Itoyama$^{a, b,c}$\footnote{e-mail: itoyama@sci.osaka-cu.ac.jp},
  Sota Nakajima$^b$\footnote{e-mail: sotanaka@sci.osaka-cu.ac.jp}   }\\

%
\vspace{10pt}
%

$^a$\it Nambu Yoichiro Institute of Theoretical and Experimental Physics (NITEP),\\
Osaka City University\\
\vspace{5pt}

$^b$\it Department of Mathematics and Physics, Graduate School of Science,\\
Osaka City University\\
\vspace{5pt}

$^c$\it Osaka City University Advanced Mathematical Institute (OCAMI)

\vspace{5pt}

3-3-138, Sugimoto, Sumiyoshi-ku, Osaka, 558-8585, Japan \\

\end{center}
%
\vspace{15pt}
\begin{center}
Abstract\\
\end{center}
We investigate the structure of the moduli space of 9D heterotic interpolating models with a complete set of Wilson line backgrounds and a radius parameter by computing the one-loop partition functions and the cosmological constants and by deriving the massless spectra, paying attention to the region $a \approx 0$ where supersymmetry is asymptotically restoring.
We find some special planes and points in the moduli space where the gauge symmetry is enhanced and/or massless fermions appear. The gauge symmetry is maximally enhanced at the minima of the one-loop effective potential where the cosmological constant is negative.
 

\vfill

\end{titlepage}

\renewcommand{\thefootnote}{\arabic{footnote}}
\setcounter{footnote}{0}


\section{Introduction}
Evidence for supersymmetry in accessible energy scale has not been found according to the LHC experiment. Also, it is not a unique possibility that we start from supersymmetric string models in 10 dimensions\cite{Gross:1984dd} and to compactify the models, preserving supersymmetry\cite{Candelas:1985en,Strings on Orbifolds,Strings on Orbifolds. 2.}. The landscape of non-supersymmetric string theories is larger than that of supersymmetric ones\cite{Kawai:1986ah,Kawai:1986va,Lerche:1986cx}, and non-supersymmetric string models can be regarded as good starting points in string phenomenology. 
{The exploration of the landscape of heterotic string vacua has been developed in Ref. \cite{Dienes:2006ut,Dienes:2007ms,Dienes:2008tg,Dienes:2008rm}.
It is worth trying to construct realistic models such as the standard(-like) model or non-supersymmetric grand unified theories (GUT) from non-supersymmetric string models. In particular, the $SO(16)\times SO(16)$ heterotic model\cite{Dixon:1986iz,AlvarezGaume:1986jb,Nair:1986zn,Ginsparg:1986wr,Itoyama:1986ei,Itoyama:1987rc,Itoyama:2019yst}, in which a tachyonic state does not appear, is often focused in the investigation of non-supersymmetric string phenomenology\cite{Blaszczyk:2014qoa,Nibbelink:2015vha,Nibbelink:2015ena,Blaszczyk:2015zta,Ashfaque:2015vta,Hamada:2015ria,McGuigan:2019gdb}. In general, however, the cosmological constants (vacuum energies) of non-supersymmetric strings are nonzero as the cancellation caused by fermion-boson degeneracy does not exist. Consequently, the dilaton tadpoles, which are proportional to the cosmological constants, are non-vanishing and cause vacuum instabilities. In order to improve this serious problem, the cosmological constant needs to be made sufficiently small. While some constructions of non-supersymmetric string models with small or zero cosmological constants have been proposed\cite{ATKIN-LEHNER SYMMETRY,Model Building on Asymmetric Z(3) Orbifolds: Nonsupersymmetric Models,The Failure of Atkin-lehner Symmetry for Lattice Compactified Strings,GENERALIZED ATKIN-LEHNER SYMMETRY,Vacuum energy cancellation in a nonsupersymmetric string,On vanishing two loop cosmological constants in nonsupersymmetric strings,Non-supersymmetric Asymmetric Orbifolds with Vanishing Cosmological Constant,More on Non-supersymmetric Asymmetric Orbifolds with Vanishing Cosmological Constant,GrootNibbelink:2017luf}, we in this paper focus on the interpolating models\cite{Itoyama:1986ei}, which are constructed by the Scherk-Schwartz compactification or the coordinate dependent compactification (CDC)\cite{Spontaneous Breaking of Supersymmetry Through Dimensional Reduction,Spontaneous Supersymmetry Breaking in Supersymmetric String Theories,Kounnas:1989dk}. The interpolating models are frequently adopted in the construction of realistic models\cite{Faraggi:2009xy,Abel:2015oxa,Aaronson:2016kjm,Abel:2017vos} or the exploration of S-duality between non-supersymmetric string theories\cite{Blum:1997gw,Blum:1997cs}. The generic partition function of 9D interpolating models is written as
\begin{align}
\label{Z_int}
Z^{(9)}_{\text{int}}
&=\frac{1}{2}Z^{(7)}_{B}\left\lbrace \Lambda_{0,0}\left(Z^{+}_{+}+Z^{+}_{-}\right) +\Lambda_{1/2,0}\left(Z^{+}_{+}-Z^{+}_{-}\right)\right. \nonumber\\ 
&\left. ~~~~~~~~+\Lambda_{0,1/2}\left(Z^{-}_{+}+Z^{-}_{-}\right)+\Lambda_{1/2,1/2}\left(Z^{-}_{+}-Z^{-}_{-}\right) \right\rbrace,
\end{align}
where $Z_{B}^{(N)}=\tau_{2}^{-N/2}\left( \eta \bar{\eta} \right)^{-N} $ is the contributions from bosonic string coordinates in spacetime directions, $Z_{\pm}^{\pm}$ represents those from fermionic string coordinates and rank 16 current algebras\footnote{The superscript and subscript of $Z_{\pm}^{\pm}$ represent the periodicities around $\sigma$ and $\tau$-directions of the world-sheet torus respectively.}, and $\Lambda_{\alpha,\beta}$ is the $SO(1,1)$ momentum lattice defined as
\begin{equation}\label{lattices}
\Lambda_{\alpha,\beta} = (\eta \bar{\eta})^{-1} \sum_{n\in 2(\boldsymbol{Z}+\alpha)} \sum_{w\in\boldsymbol{Z}+\beta}q^{\frac{\alpha'}{2} p_{L}^2} \bar{q}^{\frac{\alpha'}{2} p_{R}^2}= (\eta \bar{\eta})^{-1} \sum_{n\in 2(\boldsymbol{Z}+\alpha)} \sum_{w\in\boldsymbol{Z}+\beta}q^{\frac{1}{4} \left(na+w/a \right) ^2} \bar{q}^{\frac{1}{4}\left(na-w/a \right) ^2},
\end{equation}
with a dimensionless inverse radius $a=\sqrt{\alpha'}/R$. In the limits that $a \to 0$ and $a \to \infty$, the behaviors of the partition function \eqref{Z_int} are respectively 
\begin{align}
Z^{(9)}_{\text{int}} \underset{a\to 0}{\to} \frac{1}{2a} Z_{B}^{(8)} Z^{+}_{+},~~~Z^{(9)}_{\text{int}} \underset{a\to \infty}{\to} \frac{a}{2} Z_{B}^{(8)} \left( Z^{+}_{+}+Z^{+}_{-}+Z^{-}_{+}+Z^{-}_{-}\right) .
\end{align}
Therefore, these two 10D endpoint models are related by using the $\boldsymbol{Z}_{2}$ action $Q:Z_{+}^{\pm}\to Z_{-}^{\pm}$ and $S$ transformation $Z^{+}_{-}\to Z^{-}_{+}$ discretely\footnote{In Appendix \ref{endpoints}, some concrete examples of $Z_{+}^{+}$ and the $\boldsymbol{Z}_{2}$ action $Q$ are provided.}, while they can be connected continuously through the partition function \eqref{Z_int} with a continuous parameter. In this sense, the 9D string model with the partition function \eqref{Z_int} interpolates between the different 10D string models. In particular, if we choose $Z_{+}^{+}$ as a supersymmetric one, the cosmological constant at one-loop level can be written as\cite{Itoyama:1986ei,Itoyama:1987rc}
\begin{equation}\label{cc}
\Lambda^{(9)} = (n_F-n_B)\xi a^{9} + \mathcal{O}(e^{-1/a}),~~~a\approx 0,
\end{equation}
where $\xi $ is a computable positive constant, and $n_F$ and $n_B$ represent the degrees of freedom of massless fermions and massless bosons respectively. This formula of the cosmological constant is reviewed in Appendix \ref{calculation of cc}. The interpolating model with $n_F=n_B$ has an exponentially suppressed cosmological constant in the region $a\approx 0$ where supersymmetry is asymptotically restoring. Note that mass splitting of the broken supermultiplet is given by $\alpha' M_{s}^{2}=a^{2}$ Such models, in which the suppression of the cosmological constant occurs at one-loop level, are referred to as super no-scale models and studied in order to respect the scenario that supersymmetry is already broken at string scale\cite{Kounnas:2015yrc,Kounnas:2016gmz,Kounnas:2017mad,Coudarchet:2017pie,Coudarchet:2018ztz,Partouche:2018ftj,Florakis:2016ani,Abel:2017rch}.

Interpolating models can be deformed by turning on constant backgrounds as they are constructed by compactifying higher-dimensional string models. In heterotic strings with $d$ dimensions compactified, we have the freedom of such deformations which are represented by the coset
\begin{align}\label{coset}
\frac{SO(16+d,d)}{SO(16+d)\times SO(d)},
\end{align}
where the denominator $SO(16+d)\times SO(d)$ implies the rotations that leave left- and right-moving momenta invariant respectively. These deformations can be realized by adding the following terms to the world-sheet actions\cite{New Heterotic String Theories in Uncompactified Dimensions $<$ 10,Narain:1986am}:
\begin{align}\label{constant background}
A_{Ii}\int d^2 z \partial X^{I}_L \bar{\partial} X^{i}_R + C_{ji}\int d^2 z \partial X^{j}_L \bar{\partial} X^{i}_R,
\end{align}
where the constant $A_{Ii}$ corresponds to the background gauge fields (Wilson lines) and the constant $C_{ji}$ is decomposed into metric $g_{ji}$ and antisymmetric tensor $B_{ji}$ for $I=1,\cdots,16$, $i,j=10-d,\cdots,9$. In fact, it can be checked that the degrees of freedom of the coset \eqref{coset} agree with those of the constant backgrounds $A_{Ii}$ and $C_{ji}$. At generic points in moduli space, the gauge symmetry is broken down to the product of Abelian groups since no massless vectors are sitting on the nonzero roots in the momentum lattice. There are, however, special planes and points in moduli space where the gauge symmetry is enhanced. For example, in bosonic strings compactified on a circle, the gauge symmetry is $U(1)^{2}$ for generic radii. However, at the point in the moduli space such that the radius is $\sqrt{\alpha'}$, the gauge symmetry is enhanced to $SU(2)\times SU(2)$. In this paper, we consider 9D heterotic interpolating models and the moduli space is 17-dimensional: a radius $R$ and sixteen Wilson lines $A^{I}$. On the other hand, due to supersymmetry breaking, some moduli have no longer flat directions and should be determined. Namely, there are stable points in moduli space which correspond to the minima of the effective potential.

In Ref. \cite{Itoyama:2019yst}, we constructed 9D interpolating models with two moduli by setting $A^{1}=A,~A^{I\neq 1}=0$ and found some special points in the two-dimensional moduli space, which is reviewed in Sect. \ref{one WL}. In Sect. \ref{General 9D interpolating model}, we construct more general 9D interpolating models with a complete set of Wilson lines and find some special planes and points in the moduli space where the additional massless states appear. Sect. \ref{conclusion} is devoted to conclusion and discussion.

In Appendix \ref{notations}, we list the notation for some functions used in this paper. In Appendix \ref{endpoints}, some 10D string endpoint models and the relations between them are briefly reviewed. In Appendix \ref{calculation of cc}, we review the leading behavior of the cosmological constant of the interpolating models in the region where supersymmetry is asymptotically restoring. In Appendix \ref{appendix_momentum lattice}, we expand the momentum lattices with Lorentzian signature $(17,1)$, which appear in the partition functions with sixteen Wilson lines. These expansions are useful to identify the spectra of the string models.




\section{The review of the one Wilson line case}\label{one WL}

In this section, we briefly review Ref. \cite{Itoyama:2019yst}, in which we focus on the two-dimensional $R$-$A$ subspace of the moduli space. We evaluate the cosmological constant of the models and discuss the stability of the Wilson line.

Turning on one Wilson line background, the left- and right-moving internal momenta $\ell_{L}(=\ell^{I=1}_{L})$, $p_{L}(=p^{i=9}_{L})$ and $p_{R}(=p^{i=9}_{R})$ are changed as follows\cite{New Heterotic String Theories in Uncompactified Dimensions $<$ 10,Narain:1986am}:
\begin{eqnarray}\label{changed momenta1}
\left\{
\begin{array}{l}
\ell_{L}=\frac{1}{\sqrt{\alpha'}}m\\
p_{L}=\frac{1}{\sqrt{2\alpha'}}\left( an+a^{-1}w\right)\\
p_{R}=\frac{1}{\sqrt{2\alpha'}}\left( an-a^{-1}w\right)
\end{array}
\right.
\to
\left\{
\begin{array}{l}
\ell'_{L}=\frac{1}{\sqrt{\alpha'}}\left( m-2t_1 w \right) \\
p'_{L}=\frac{1}{\sqrt{\alpha'}}t_{2}^{-1}\left( t_1m+n/2-(t_{1}^2-t_{2}^2)w \right)\\
p'_{R}=\frac{1}{\sqrt{\alpha'}}t_{2}^{-1}\left( t_1m+n/2-(t_{1}^2+t_{2}^2)w \right),
\end{array}
\right.
\end{eqnarray}
where $t$ is defined as
\begin{equation}
t=t_1+it_2=\frac{1}{\sqrt{2}}Aa_{0}^{-1}+\frac{i}{\sqrt{2}} a_{0}^{-1},
\end{equation}
for $ a_0\equiv\sqrt{1+A^2}a$.
In our partition function, a theta function which represents the sum over the zero modes of $X_{L}^{I=1}$ is combined with the momentum lattice $\Lambda_{\alpha,\beta}$ as follows:
\begin{equation}\label{changed point}
\Lambda_{\alpha,\beta}\ \vartheta
	\begin{bmatrix} 
	\gamma\\ 
	\delta\\ 
	\end{bmatrix}\eta^{-1}\to
\Lambda^{(\alpha,\beta)}_{(\gamma,\delta)}(a,A),
\end{equation}
where $\Lambda^{(\alpha,\beta)}_{(\gamma,\delta)}$ is defined as
\begin{equation}
\label{mixed lattice}
\Lambda^{(\alpha,\beta)}_{(\gamma,\delta)}(a,A) \equiv \left( \eta \bar{\eta}\right)^{-1} \eta^{-1} \sum_{n,w,m} (-1)^{2m\delta} q^{\frac{\alpha'}{2}\left( p'^{2}_{L}+\ell'^{2}_{L}\right) } \bar{q}^{\frac{\alpha'}{2} p'^{2}_{R}},
\end{equation}
for $n\in2(\boldsymbol{Z}+\alpha)$, $w\in\boldsymbol{Z}+\beta$, $m\in\boldsymbol{Z}+\gamma$. We can check that the lattice $\Lambda^{(\alpha,\beta)}_{(\gamma,\delta)}$ is invariant under the shift
\begin{equation}\label{shift symm}
t\to t+2.
\end{equation}
So, the fundamental region of the moduli space is
\begin{equation}
\label{fund region}
-1<t_{1}\leq1,~~t_{2}\geq 0.
\end{equation}

\subsection{Two examples}

Let us construct concrete examples of 9D interpolating models with one Wilson line. We choose the $SO(16)\times SO(16)$ model as the 10D non-supersymmetric endpoint and consider the following two interpolations:
\begin{itemize}
	\item Model I: 10D SUSY $SO(32)$ model $\longleftrightarrow$ 10D $SO(16)\times SO(16)$ model
	\item Model I\hspace{-.1em}I: 10D SUSY $E_8 \times E_8$ model $\longleftrightarrow$ 10D $SO(16)\times SO(16)$ model
\end{itemize}
Applying \eqref{changed point} to \eqref{Z_int}, we can obtain the partition functions of Model I and Model I\hspace{-.1em}I which are respectively written as follows:
\begin{align}
\label{model A one WL}
Z^{(9)}_{\text{Model I}}
&= Z^{(7)}_{B}\left\lbrace \bar{V}_{8}\left( O^{(0,0)}_{16}O_{16}+S^{(0,0)}_{16}S_{16}\right) -\bar{S}_{8}\left( V^{(0,0)}_{16}V_{16}+C^{(0,0)}_{16}C_{16}\right)\right. \nonumber\\
&~~~~~\left. +\bar{V}_{8}\left( V^{(1/2,0)}_{16}V_{16}+C^{(1/2,0)}_{16}C_{16}\right) -\bar{S}_{8}\left( O^{(1/2,0)}_{16}O_{16}+S^{(1/2,0)}_{16}S_{16}\right)\right.\nonumber\\
&~~~~~\left. +\bar{O}_{8}\left( V^{(0,1/2)}_{16}C_{16}+C^{(0,1/2)}_{16}V_{16}\right) -\bar{C}_{8}\left( O^{(0,1/2)}_{16}S_{16}+S^{(0,1/2)}_{16}O_{16}\right)\right.\nonumber\\
&~~~~~\left. +\bar{O}_{8}\left( O^{(1/2,1/2)}_{16}S_{16}+S^{(1/2,1/2)}_{16}O_{16}\right) -\bar{C}_{8}\left( V^{(1/2,1/2)}_{16}C_{16}+C^{(1/2,1/2)}_{16}V_{16}\right)\right\rbrace, \\
\label{model B one WL}
Z^{(9)}_{\text{Model I\hspace{-.1em}I}}
&= Z^{(7)}_{B}\left\lbrace \bar{V}_{8}\left( O^{(0,0)}_{16}O_{16}+S^{(0,0)}_{16}S_{16}\right) -\bar{S}_{8}\left( O^{(0,0)}_{16}S_{16}+S^{(0,0)}_{16}O_{16}\right)\right. \nonumber\\
&~~~~~\left. +\bar{V}_{8}\left( O^{(1/2,0)}_{16}S_{16}+S^{(1/2,0)}_{16}O_{16}\right) -\bar{S}_{8}\left( O^{(1/2,0)}_{16}O_{16}+S^{(1/2,0)}_{16}S_{16}\right)\right.\nonumber\\
&~~~~~\left. +\bar{O}_{8}\left( V^{(0,1/2)}_{16}C_{16}+C^{(0,1/2)}_{16}V_{16}\right) -\bar{C}_{8}\left( V^{(0,1/2)}_{16}V_{16}+C^{(0,1/2)}_{16}C_{16}\right)\right.\nonumber\\
&~~~~~\left. +\bar{O}_{8}\left( V^{(1/2,1/2)}_{16}V_{16}+C^{(1/2,1/2)}_{16}C_{16}\right) -\bar{C}_{8}\left( V^{(1/2,1/2)}_{16}C_{16}+C^{(1/2,1/2)}_{16}V_{16}\right)\right\rbrace,
\end{align}
where we define $\left( O^{(\alpha,\beta)}_{2n}, V^{(\alpha,\beta)}_{2n}, S^{(\alpha,\beta)}_{2n}, C^{(\alpha,\beta)}_{2n}\right)$ as
\begin{align}
\left(
\begin{array}{c}
O^{(\alpha,\beta)}_{2n}\\
V^{(\alpha,\beta)}_{2n}
\end{array}
\right)&\equiv\frac{1}{2\eta^{n-1}}\left(  \Lambda^{(\alpha,\beta)}_{(0,0)}\vartheta^{n-1}
\begin{bmatrix} 
0\\ 
0\\ 
\end{bmatrix}(0,\tau)\pm  \Lambda^{(\alpha,\beta)}_{(0,1/2)}\vartheta^{n-1}
\begin{bmatrix} 
0\\ 
1/2\\ 
\end{bmatrix}(0,\tau)
\right),\\
\left(
\begin{array}{c}
S^{(\alpha,\beta)}_{2n}\\
C^{(\alpha,\beta)}_{2n}
\end{array}
\right)&\equiv\frac{1}{2\eta^{n-1}}\left( \Lambda^{(\alpha,\beta)}_{(1/2,0)}\vartheta^{n-1}
\begin{bmatrix} 
1/2\\ 
0\\ 
\end{bmatrix}(0,\tau)\pm \Lambda^{(\alpha,\beta)}_{(1/2,1/2)}\vartheta^{n-1}
\begin{bmatrix} 
1/2\\ 
1/2\\ 
\end{bmatrix}(0,\tau)
\right).
\end{align}
Note that $Z^{(9)}_{\text{Model I}}$ and $Z^{(9)}_{\text{Model I\hspace{-.1em}I}}$ depend on $a$ and $A$ through $ \Lambda^{\left( \alpha,\beta\right)}_{\left( \gamma,\delta\right)}\left(a,A \right) $.
We can see the spectrum of the model at each mass level by expanding the partition function in $q$. At generic values of $t$, the massless spectrum of Model I is
\begin{itemize}
	\item the nine-dimensional gravity multiplet: graviton $G_{\mu \nu}$, anti-symmetric tensor $B_{\mu \nu}$ and dilaton $\phi$;
	\item the gauge bosons of $SO(16)\times SO(14)\times U(1)^3$;
	\item a spinor transforming in the $(\boldsymbol{16},\boldsymbol{14})$ of $SO(16)\times SO(14)$.
\end{itemize}
In Model I\hspace{-.1em}I, the massless bosons are the same as in Model I while the massless spinors transform in the $(\boldsymbol{128},\boldsymbol{1})$ of $SO(16)\times SO(14)$.
Note that the Abelian factors of the gauge group contain $U(1)^2$ generated by $G_{\mu 9}$ and $B_{\mu 9}$. As expected, there are special points in the moduli space where the additional states become massless.
We summarize the special points in the moduli space and the additional massless states at the points in Tabel 1 and Table 2. Although only the non-Abelian parts of gauge symmetry are shown in the tables, there is, in fact, the product of $U(1)$'s so that the total rank of the group is 18\footnote{In this paper, we sometimes omit the Abelian factors. But, the total rank of the gauge symmetry is always 18 as long as we consider continuous deformations such as \eqref{constant background}.}. These tables show only the special points at which all the massless states have zero winding numbers because we are interested in the region $a\approx 0$ where supersymmetry is asymptotically restoring. In Model I, the cosmological constant is exponentially suppressed at the points with $t_1=\pm1/2$.

\begin{table}[t]
	\centering
	\begin{tabular}{|c||c|c|c|} \hline
		Special points & $t_1=0,1$  & $t_1=\pm 1/2$\\ \hline 
		Gauge symmetry & $SO(16)\times SO(16)$& $SO(18) \times SO(14)$   \\ \hline
		Massless spinors & $(\boldsymbol{16},\boldsymbol{16})$& $(\boldsymbol{18},\boldsymbol{14})$ \\ \hline
		$n_F-n_B$& positive & zero  \\ \hline
	\end{tabular}
	\caption{The special points and the massless states at the points in Model I. } 
	\vspace{20pt}
	\begin{tabular}{|c||c|c|c|} \hline
		Special points & $t_1=0$& $t_1=1$ & $t_1=\pm 1/2$ \\ \hline 
		Gauge symmetry & $SO(16)\times SO(16)$& $SO(16) \times E_8$   &$SO(16)\times SO(14)$\\ \hline
		Massless spinors & $(\boldsymbol{128},\boldsymbol{1})\oplus(\boldsymbol{1},\boldsymbol{128})$& $(\boldsymbol{128},\boldsymbol{1})$&$(\boldsymbol{128},\boldsymbol{1})$\\ \hline
		$n_F-n_B$& positive & negative  & negative\\ \hline
	\end{tabular}
	\caption{The special points and the massless states at the points in Model I\hspace{-.1em}I.}
\end{table}

\subsection{The cosmological constant}\label{cc_oneWL}

We can evaluate the cosmological constant in the region $a\approx 0$, following the procedure in Appendix \ref{calculation of cc}. 
The cosmological constants of Model I and Model I\hspace{-.1em}I are respectively
\begin{align}
\Lambda_{\text{Model I}}^{(9)}(a,A) 
&\simeq C_{0} \left( \frac{a_{0}}{\sqrt{\alpha'}} \right)^9 8\left\lbrace (224-220) + 2(16-14) \cos\left( 2\pi t_1\right)  \right\rbrace,\\
\Lambda_{\text{Model I\hspace{-.1em}I}}^{(9)}(a,A) 
&\simeq C_{0} \left( \frac{a_{0}}{\sqrt{\alpha'}} \right)^9 8\left\lbrace (2^7-220) - 2\cdot 14 \cos\left( 2\pi t_1\right)  + 2\cdot  2^6\cos\left(\pi t_1 \right) \right\rbrace,
\end{align}
where $C_{0}$ is a positive constant.
Note that in Model I, the cosmological constant is already invariant under the shift $t_1 \to t_1+1$. So, in Model I, we can restrict our attention to the region $-1/2<t_{1}\leq1/2$ when $a\approx 0$.

\begin{figure}[t]
	\begin{minipage}{0.5\hsize}
		\begin{center}
			\includegraphics[width=75mm]{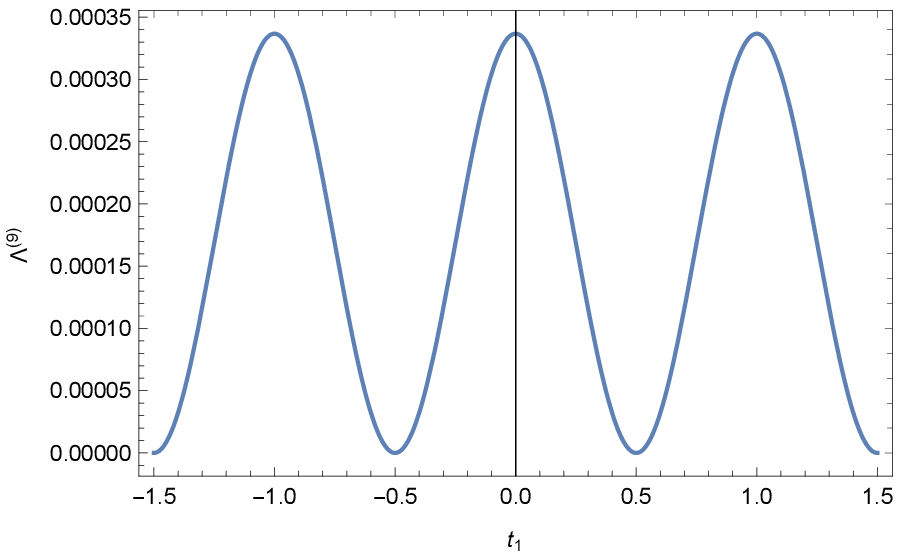}
		\end{center}
		\label{fig:one}
	\end{minipage}
	\begin{minipage}{0.5\hsize}
		\begin{center}
			\includegraphics[width=75mm]{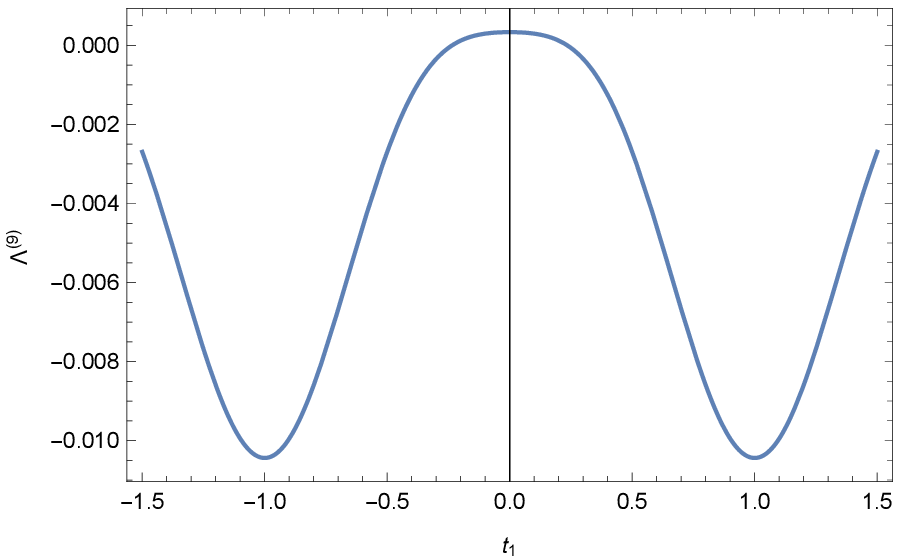}
		\end{center}
		\label{fig:two}
	\end{minipage}
	\caption{The left and right figures show the cosmological constants $\Lambda^{(9)}$ with $a_0=1$ of Model I and Model I\hspace{-.1em}I respectively, in units of $\alpha'=1$, up to exponentially suppressed terms. The cosmological constants decay albegraically as $t_{2}\to \infty$. }
\end{figure}

Fig. 1 shows the cosmological constants of Model I and I\hspace{-.1em}I in terms of $t_1$ respectively. From Tables 1 and 2, we find that the gauge symmetry is enhanced at the extrema of $\Lambda^{(9)}$. In particular, it seems that in Model I, the minima of $\Lambda^{(9)}$ are the points corresponding to the $SO(18)\times SO(14)$ enhanced gauge symmetry, where the cosmological constant is exponentially suppressed. In the next section, however, we will see that the other Wilson lines are unstable at these points.

\section{General 9D interpolating models}\label{General 9D interpolating model}

In this section, we generalize 9D interpolating models by turning on a complete set of Wilson line backgrounds. In one-dimensional compactification of heterotic strings with sixteen Wilson lines, the internal momenta are given as follows\cite{New Heterotic String Theories in Uncompactified Dimensions $<$ 10,Narain:1986am}:
\begin{align}
\ell_{L}^I&=\frac{1}{\sqrt{2\alpha'}}\left( \sqrt{2}m^I-2A^I a_{0}^{-1} w \right), \\
p_{L}&=\frac{1}{\sqrt{2\alpha'}}\left( \sqrt{2} A\cdot m+a_{0}n-(1-|A|^2)a_{0}^{-1}w \right),\\
p_{R}&=\frac{1}{\sqrt{2\alpha'}}\left( \sqrt{2} A\cdot m+a_{0}n-(1+|A|^2)a_{0}^{-1}w \right),
\end{align}
where $A\cdot m\equiv\sum_{I=1}^{16}A^{I}m^{I}$, $|A|^2\equiv A\cdot A$, and $a_{0}\equiv\sqrt{1+|A|^2}a$. Defining $t_{1}^{I}$ and $t_{2}$ as
\begin{align}
t_{1}^{I} = \frac{1}{\sqrt{2}} A^I a_{0}^{-1},~~ t_{2} = \frac{1}{\sqrt{2}}a_{0}^{-1},
\end{align}
the momenta can be rewritten as
\begin{align}
\ell_{L}^I&=\frac{1}{\sqrt{\alpha'}}\left(m^I-2t_{1}^I w \right), \\
p_{L}&=\frac{1}{\sqrt{\alpha'}}t_{2}^{-1}\left( t_{1}\cdot m+\frac{n}{2}-(|t_{1}|^2-t_{2}^2)w \right),\\
p_{R}&=\frac{1}{\sqrt{\alpha'}}t_{2}^{-1}\left( t_{1}\cdot m+\frac{n}{2}-(|t_{1}|^2+t_{2}^2)w \right).
\end{align}

In order to obtain the partition function, we define a momentum lattice $\Lambda^{(\alpha,\beta)}_{(\gamma^I,\delta^I)}$ with Lorentzian signature $(17,1)$ as
\begin{align}
\Lambda^{(\alpha,\beta)}_{(\gamma^I,\delta^I)}\left(a,A^I \right) \equiv(\eta \bar{\eta})^{-1} \eta^{-16} \sum_{n,w,m^I} (-1)^{2\delta\cdot m} q^{\frac{\alpha'}{2}\left( p_{L}^2+|\ell_{L}|^2 \right) } \bar{q}^{\frac{\alpha'}{2}p_{R}^2} ,
\end{align}
where the sum is taken over $m^I \in \boldsymbol{Z}+\gamma^I$, $n\in 2(\boldsymbol{Z}+\alpha)$ and $w\in \boldsymbol{Z}+\beta$.
We find that $\Lambda^{(\alpha,\beta)}_{(\gamma^I,\delta^I)}$ is invariant under the shift 
\begin{align}\label{shifsym}
t_{1}^i \to t_{1}^i + 2 ,~~~~~\text{for any $i$,}
\end{align}
with the redefinitions
\begin{equation}
\begin{split}
m^i& \to m'^i =m^i-4w,~~~~~~m^{I\neq i} \to m'^{I\neq i}=m^{I\neq i}, \\
n &\to n'=n -8w + 4\sum_{i} m,~~~~~~w \to w'=w.
\end{split}
\end{equation}
Thus, the fundamental region of the moduli space is 
\begin{align}
-1 < t_{1}^{I} \leq 1,~~~t_{2}\geq 0
\end{align}

After sixteen Wilson line backgrounds are turned on, the effective change in the partition function is\footnote{In this paper, we focus on the interpolations between 10D string models whose left-moving parts of the partition function can be written in terms of the $SO(16)$ characters.} 
\begin{align}\label{changed point2}
\Lambda_{\alpha,\beta}\ \vartheta^8
\begin{bmatrix} 
\gamma_1\\ 
\delta_1\\ 
\end{bmatrix}\vartheta^8
\begin{bmatrix} 
\gamma_2\\ 
\delta_2\\ 
\end{bmatrix}
\eta^{-16}\to
\Lambda^{(\alpha,\beta)}\begin{bmatrix} 
\gamma_1&\gamma_2\\ 
\delta_1&\delta_2\\ 
\end{bmatrix}(a,A^I),
\end{align}
where we define
\begin{align}\label{boosted lattices}
\Lambda^{(\alpha,\beta)}\begin{bmatrix} 
\gamma_1&\gamma_2\\ 
\delta_1&\delta_2\\ 
\end{bmatrix}(a,A^I)\equiv 
\left. \Lambda^{(\alpha,\beta)}_{(\gamma^I,\delta^I)}\left(a,A^I \right)\right| _{\gamma^I=\left( (\gamma_{1})^8,(\gamma_{2})^8\right),~\delta^I=\left( (\delta_{1})^8,(\delta_{2})^8\right) } .
\end{align}
Note that the sums over $n$, $w$ and $m^{I}$ can no longer be carried out separately in the partition function.

\subsection{Interpolation between SUSY $SO(32)$ and $SO(16)\times SO(16)$}
Let us consider Model I with sixteen Wilson lines. According to \eqref{changed point2}, the partition function of Model I is written as
\begin{align}\label{modelA_16WL}
Z^{(9)}(a,A^I)
&= Z^{(7)}_{B}\left\lbrace \bar{V}_{8}\left(\chi_{OO}^{(0,0)}+\chi_{SS}^{(0,0)}\right) -\bar{S}_{8}\left( \chi_{VV}^{(0,0)}+\chi_{CC}^{(0,0)}\right)\right. \nonumber\\
&~~~~~\left. +\bar{V}_{8}\left( \chi_{VV}^{(1/2,0)}+\chi_{CC}^{(1/2,0)}\right) -\bar{S}_{8}\left( \chi_{OO}^{(1/2,0)}+\chi_{SS}^{(1/2,0)}\right)\right.\nonumber\\
&~~~~~\left. +\bar{O}_{8}\left( \chi_{VC}^{(0,1/2)}+ \chi_{CV}^{(0,1/2)}\right) -\bar{C}_{8}\left( \chi_{OS}^{(0,1/2)}+\chi_{SO}^{(0,1/2)}\right)\right.\nonumber\\
&~~~~~\left. +\bar{O}_{8}\left( \chi_{OS}^{(1/2,1/2)}+\chi_{SO}^{(1/2,1/2)}\right) -\bar{C}_{8}\left( \chi_{VC}^{(1/2,1/2)}+\chi_{CV}^{(1/2,1/2)}\right)\right\rbrace, 
\end{align}
where we define
\begin{equation}
\begin{split}
\chi_{OO}^{(\alpha,\beta)}&\equiv \frac{1}{4} \sum_{\delta_{1},\delta_{2}=0,1/2} \Lambda^{(\alpha,\beta)}\begin{bmatrix} 
0&0\\ 
\delta_1&\delta_2\\ 
\end{bmatrix},~~~~~~~
\chi_{VV}^{(\alpha,\beta)}\equiv \frac{1}{4} \sum_{\delta_{1},\delta_{2}=0,1/2} e^{2\pi i (\delta_{1}+\delta_{2})}
\Lambda^{(\alpha,\beta)}\begin{bmatrix} 
0&0\\ 
\delta_1&\delta_2\\ 
\end{bmatrix},\\
\chi_{SS}^{(\alpha,\beta)}&\equiv \frac{1}{4} \sum_{\delta_{1},\delta_{2}=0,1/2} \Lambda^{(\alpha,\beta)}\begin{bmatrix} 
1/2&1/2\\ 
\delta_1&\delta_2\\ 
\end{bmatrix},~~~~~
\chi_{CC}^{(\alpha,\beta)}\equiv \frac{1}{4} \sum_{\delta_{1},\delta_{2}=0,1/2} e^{2\pi i (\delta_{1}+\delta_{2})}
\Lambda^{(\alpha,\beta)}\begin{bmatrix} 
1/2&1/2\\ 
\delta_1&\delta_2\\ 
\end{bmatrix},\\
\chi_{OS}^{(\alpha,\beta)}&\equiv \frac{1}{4} \sum_{\delta_{1},\delta_{2}=0,1/2} \Lambda^{(\alpha,\beta)}\begin{bmatrix} 
0&1/2\\ 
\delta_1&\delta_2\\ 
\end{bmatrix},~~~~~~~
\chi_{VC}^{(\alpha,\beta)}\equiv \frac{1}{4} \sum_{\delta_{1},\delta_{2}=0,1/2} e^{2\pi i (\delta_{1}+\delta_{2})}
\Lambda^{(\alpha,\beta)}\begin{bmatrix} 
0&1/2\\ 
\delta_1&\delta_2\\ 
\end{bmatrix},\\
\chi_{SO}^{(\alpha,\beta)}&\equiv \frac{1}{4} \sum_{\delta_{1},\delta_{2}=0,1/2} \Lambda^{(\alpha,\beta)}\begin{bmatrix} 
1/2&0\\ 
\delta_1&\delta_2\\ 
\end{bmatrix},~~~~~~~
\chi_{CV}^{(\alpha,\beta)}\equiv \frac{1}{4} \sum_{\delta_{1},\delta_{2}=0,1/2} e^{2\pi i (\delta_{1}+\delta_{2})}
\Lambda^{(\alpha,\beta)}\begin{bmatrix} 
1/2&0\\ 
\delta_1&\delta_2\\ 
\end{bmatrix}.
\end{split}
\end{equation}
In Appendix \ref{appendix_momentum lattice}, we give some properties of $\Lambda^{(\alpha,\beta)}_{(\gamma^I,\delta^I)}$ and $\chi_{**}^{(\alpha,\beta)}$ in the region $a\approx 0$.

\subsubsection{Massless spectrum}

Let us see the massless spectrum by expanding $Z^{(9)}(a,A^I)$ in $q$. As in Sect. \ref{one WL}, we restrict our attention to the states with zero winding number, which means that the parts with $\beta=1/2$ in the partition function are omitted since we are interested in the behavior of the model at the region $a\approx 0$. In this section, we assume that $t_{2}$ is fixed and take a large value such that the formula \eqref{cc} is valid, and consider the 16-dimensional moduli space characterized by $t_{1}^{I}$.

At generic points in the moduli space, the massless states appear from $\bar{V}_8 \chi_{OO}^{(0,0)}$ only when $m^I =0$ and $n=w=0$. Thus, the massless spectrum at generic points is
\begin{itemize}
	\item the nine-dimensional gravity multiplet: $G_{\mu \nu}$, $B_{\mu \nu}$, $\phi$;
	\item the gauge bosons of $U(1)^{18}$.
\end{itemize}
That the gauge symmetry is broken down to $U(1)^{18}$ means that the momenta of the massless vectors no longer get on the points corresponding to nonzero roots of any Lie algebras in the momentum lattice, due to the deformations of the Wilson lines. As in the previous section, however, we can find some special planes and points in the moduli space where the additional massless states appear. 

Let us denote $t_{1}^{I}=\left( t_{1}^{A},t_{1}^{A'} \right)$ with $A=1,\cdots,8$, $A'=9,\cdots,16$, and consider a plane in the moduli space where $p$ of $t_{1}^{A}$'s take the same value:
\begin{align}\label{condition1}
t_{1}^{a_1}=t_{1}^{a_2}=\cdots=t_{1}^{a_p}=x,~~~~\text{for $p\geq 2$},
\end{align}
Note that the constraint \eqref{condition1} represents the line in the $p$-dimensional subspace of the moduli space. By expanding $\bar{V}_8 \chi_{OO}^{(0,0)}$ on this plane, we find the additional massless vectors with 
\begin{align}
m^{I\neq a}=0,~~~m^{a}=\left( \underline{+1, -1, (0)^{p-2}}\right),
\end{align}
where the underline represents the permutations of the components and the index $a$ is denoted as $\left\lbrace a_1, a_2, \cdots, a_p \right\rbrace$. On the plane satisfying \eqref{condition1}, therefore, the gauge symmetry is enhanced to $SU(p)\times U(1)\times U(1)^{18-p}$. Next, let us consider a more special plane on which in addition to \eqref{condition1}, the following constraint is satisfied:
\begin{align}\label{condition2}
t_{1}^{b_1}=t_{1}^{b_2}=\cdots=t_{1}^{b_q}=y,~~~~\text{for  $b\neq a$, $q\geq 2$},
\end{align}
where $1\leq b \leq8$ or $9\leq b \leq16$.
 Then, we find, from $\bar{V}_8 \chi_{OO}^{(0,0)}$, the additional massless vectors whose $\left( m^{a},m^{b}\right) $ takes the values of the nonzero roots of $SU(p)\times SU(q)$:
\begin{align}\label{su(p)su(q)}
\left( m^{a}, m^{b}\right)= \left( \underline{+1, -1, (0)^{p-2}}, (0)^{q}\right), \left( (0)^{p} ,\underline{+1, -1, (0)^{p-2}}\right),~~~~m^{I\neq a,b}=0.
\end{align}
 Furthermore, if $x+y$ or $x-y$ is an integer or a half-integer, then we find more massless vectors or/and massless spinors. Supposing $1\leq b \leq8$, the additional massless states appearing on such planes in the moduli space are as follows;
	\begin{enumerate}
		\item[(a)] $x-y \in \boldsymbol{Z}$: 
		
		$\bar{V}_8 \chi_{OO}^{(0,0)}$ has the massless vectors with
		\begin{align}\label{su(p+q)}
		\left(m^a, m^b \right) =\left( \underline{+1, -1, (0)^{p+q-2}} \right),~~m^{I\neq a,b}=0,
		\end{align}
		which correspond to the nonzero roots of $SU(p+q)$. There are no massless fermions.
		\item[(b)] $x+y \in \boldsymbol{Z}$: 
		
		The massless spectrum on these planes is the same as on the planes (a).
		
		\item[(c)] $x-y \in \boldsymbol{Z}+1/2$:
		
		$\bar{V}_8 \chi_{OO}^{(0,0)}$ has the massless vectors with \eqref{su(p)su(q)}. $\bar{S}_8 \chi_{OO}^{(1/2,0)}$ has the massless spinors with
		\begin{align}\label{(p,bar{q})}
		\left(m^a, m^b \right) =\pm\left( \underline{+1, (0)^{p-1}}, \underline{-1, (0)^{q-1}} \right),~~m^{I\neq a,b}=0,
		\end{align}
		
		which correspond to the $\left(\boldsymbol{p}, \boldsymbol{\bar{q}} \right)\oplus\left(\boldsymbol{\bar{p}}, \boldsymbol{q} \right)$ of $SU(p)\times SU(q)$.
		\item[(d)] $x+y \in \boldsymbol{Z}+1/2$:
		
		$\bar{V}_8 \chi_{OO}^{(0,0)}$ has the massless vectors with \eqref{su(p)su(q)}. $\bar{S}_8 \chi_{OO}^{(1/2,0)}$ has the massless spinors with
		\begin{align}\label{(p,q)}
		\left(m^a, m^b \right) =\pm\left( \underline{+1, (0)^{p-1}}, \underline{+1, (0)^{q-1}} \right),~~m^{I\neq a,b}=0,
		\end{align}
		which correspond to the $\left(\boldsymbol{p}, \boldsymbol{q} \right)\oplus\left(\boldsymbol{\bar{p}}, \boldsymbol{\bar{q}} \right)$ of $SU(p)\times SU(q)$.
	\end{enumerate}
	The intersections of two of the above planes are more special;
	\begin{itemize}
		\item[(a,b)]$x-y \in \boldsymbol{Z}$ and $x+y \in \boldsymbol{Z}$:
		
		$\bar{V}_8 \chi_{OO}^{(0,0)}$ has the massless vectors with
		\begin{align}\label{so(2p+2q)}
		\left(m^a, m^b \right) =\left( \underline{\pm 1, \pm 1, (0)^{p+q-2}} \right),~~m^{I\neq a,b}=0,
		\end{align}
		which correspond to the nonzero roots of $SO\left( 2(p+q) \right) $. There are no massless fermions.
		
		\item[(a,d)]$x-y \in \boldsymbol{Z}$ and $x+y \in \boldsymbol{Z}+1/2$:
		
		$\bar{V}_8 \chi_{OO}^{(0,0)}$ has the massless vectors with the same values of $\left(m^a,m^b \right) $ as in \eqref{su(p+q)}. $\bar{S}_8 \chi_{OO}^{(1/2,0)}$ has the massless spinors with
		\begin{align}\label{(antisym rep)}
		\left(m^a, m^b \right) =\pm\left( \underline{+1, +1,(0)^{p+q-2}} \right),~~m^{I\neq a,b}=0,
		\end{align}
		which correspond to the antisymmetric representation and its conjugate of $SU(p+q)$.

		\item[(b,c)]$x+y \in \boldsymbol{Z}$ and $x-y \in \boldsymbol{Z}+1/2$:
		
		The massless spectrum at these intersections is the same as at the intersections (a,d).
		\item[(c,d)]$x-y \in \boldsymbol{Z}+1/2$ and $x+y \in \boldsymbol{Z}+1/2$:
		
		$\bar{V}_8 \chi_{OO}^{(0,0)}$ has the massless vectors with
		\begin{align}\label{so(2p)so(2q)}
		\left(m^a, m^b \right) =\left( \underline{\pm 1, \pm 1, (0)^{p-2}},(0)^q \right),\left((0)^p, \underline{\pm 1, \pm 1, (0)^{q-2}} \right),~~m^{I\neq a,b}=0,
		\end{align}
		which correspond to the nonzero roots of $SO(2p)\times SO(2q)$.
		$\bar{S}_8 \chi_{OO}^{(1/2,0)}$ has the massless spinors with
		\begin{align}\label{(2p,2q)}
		\left(m^a, m^b \right) =\left( \underline{\pm 1, (0)^{p-1}}, \underline{\pm 1, (0)^{q-1}}\right)~~m^{I\neq a,b}=0,
		\end{align}
		which correspond to the nonzero roots of the $\left( \boldsymbol{2p},\boldsymbol{2q}\right)$ of $SO(2p)\times SO(2q)$.
	
	\end{itemize}
\begin{figure}[t]
	\begin{minipage}{0.5\hsize}
		\begin{center}
			\includegraphics[width=65mm]{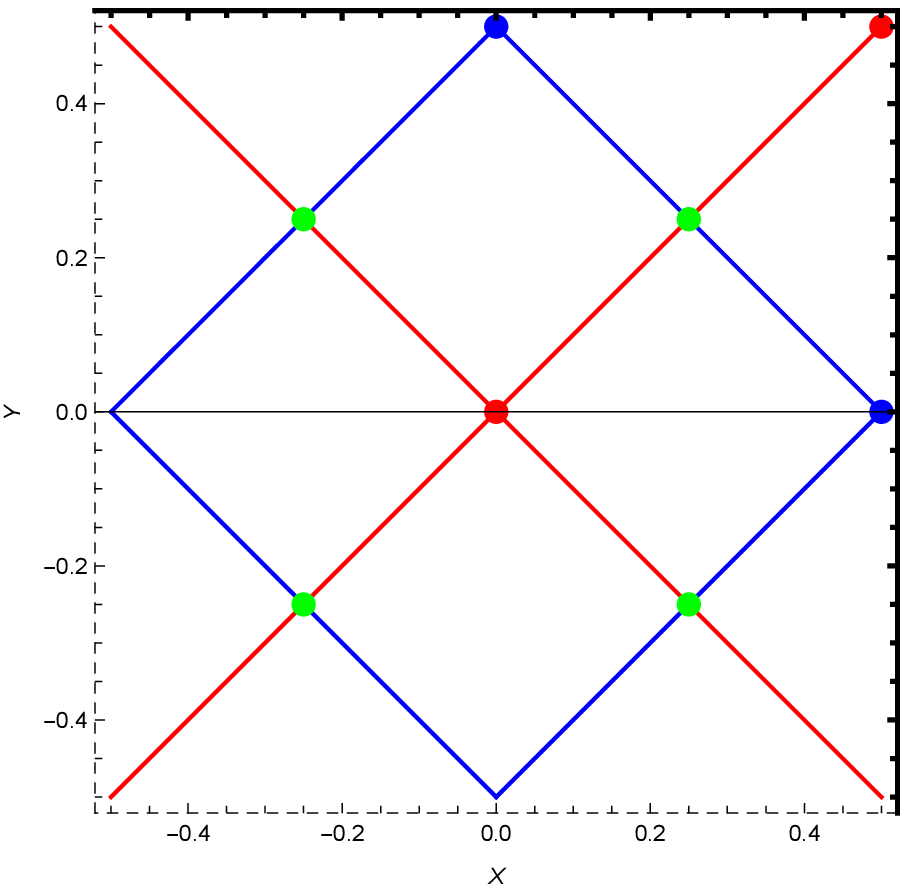}
		\end{center}
		\label{fig:moduli1}
	\end{minipage}
	\begin{minipage}{0.5\hsize}
		\begin{center}
			\includegraphics[width=65mm]{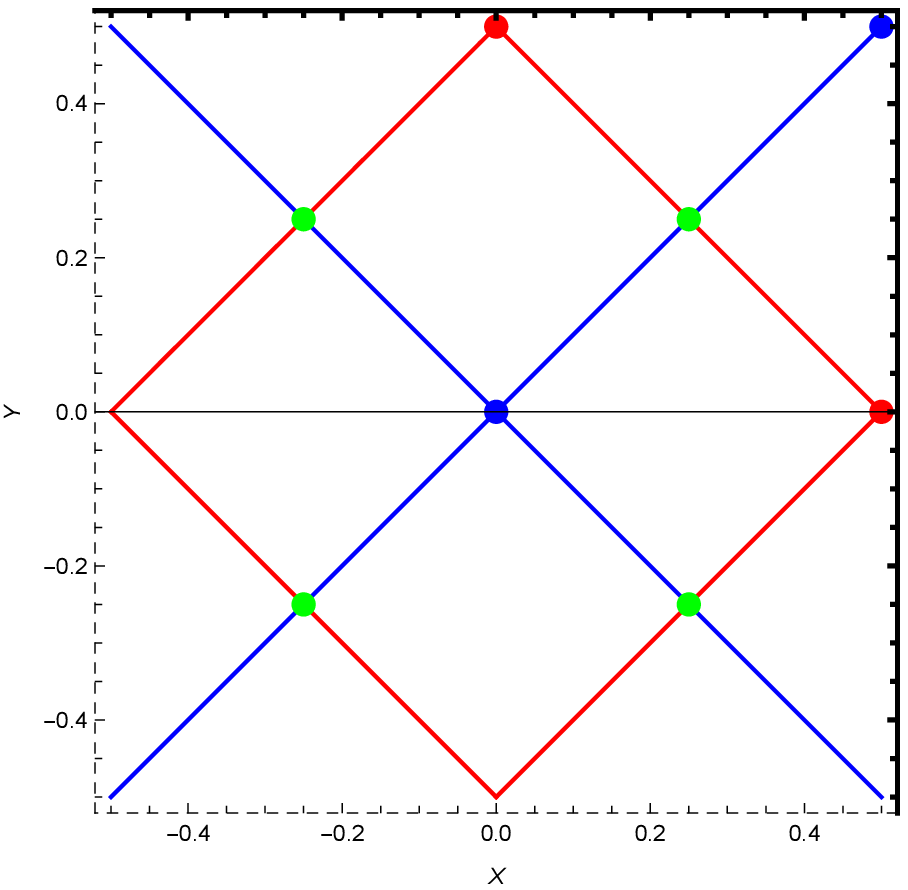}
		\end{center}
	\end{minipage}
	\caption{The left and right figures show the special planes in the moduli space in the case that $1\leq b\leq 8$ and $9\leq b\leq 16$ respectively. At generic points, the gauge symmetry is $SU(p)\times SU(q)$ and there are no massless fermions. The red and blue lines correspond to the planes (a), (b) and (c), (d) respectively. The red, blue, and green points correspond to the intersections (a,b), (a,d) or (b,c), and (a,d) respectively.}
\end{figure}
If $9\leq b \leq 16$, then the massless spectra on the planes where $x\pm y$ is an integer (a half-integer) are the same as on the planes where $x\pm y$ is a half-integer (an integer) in the $1\leq b \leq 8$ case, and the intersections (a,b) and (c,d) are exchanged accordingly. Fig. 2 shows the planes and the intersections in the fundamental region of the $x$-$y$ plane in the cases that $1\leq b\leq 8$ and $9\leq b\leq 16$. Note that the fundamental region of the moduli space becomes smaller $-1/2<t_{1}^{I}\leq 1/2$ in the region $a\approx 0$ as in the one Wilson line case, which is shown from the cosmological constant written as a function of $t_{1}^{I}$.  

 As a more special point in the moduli space, let us consider the following configuration of the Wilson lines:
\begin{align}
t_{1}^{A}=\left((0)^p, \left( \frac{1}{2}\right) ^q,  \left( \frac{1}{4}\right) ^r,  \left( -\frac{1}{4}\right) ^s \right) ,~~t_{1}^{A'}=\left((0)^{p'}, \left( \frac{1}{2}\right) ^{q'},  \left( \frac{1}{4}\right) ^{r'},  \left( -\frac{1}{4}\right) ^{s'} \right),
\end{align}
for $p+q+r+s=p'+q'+r'+s'=8$. The massless spectrum at this point is
\begin{itemize}
	\item the nine-dimensional gravity multiplet: graviton $G_{\mu \nu}$, anti-symmetric tensor $B_{\mu \nu}$ and dilaton $\phi$;
	\item the gauge bosons of $SO(2P)\times SO(2Q) \times \left( SU(R)\times U(1)\right) \times U(1)^{2}$;
	\item  the spinors transforming in $\left( \boldsymbol{2P},\boldsymbol{2Q},\boldsymbol{1}\right)\oplus \left( \boldsymbol{1},\boldsymbol{1},\boldsymbol{\frac{R(R-1)}{2}}\right) \oplus \left( \boldsymbol{1},\boldsymbol{1},\overline{\boldsymbol{\frac{R(R-1)}{2}}}\right)$ of $SO(2P)\times SO(2Q) \times SU(R)$,
\end{itemize}
where $P\equiv p+q'$, $Q\equiv q+p'$, and $R\equiv r+s+r'+s'$. 
Note that at the points with $p=q'=8$ or $q=p'=8$, the gauge symmetry is maximally enhanced to $SO(32)$ and there are no massless fermions.

\subsubsection{Cosmological constant and stability}

 Following the procedure in Appendix \ref{calculation of cc}, the cosmological constant in the region $a\approx 0$ is
\begin{align}
\Lambda^{(9)}(a,A^I) &\simeq C_{0}\left( \frac{a_0}{\sqrt{\alpha'}} \right)^9 8 \left\lbrace 
4 \sum_{A=1}^{8} \sum_{A'=9}^{16} \cos \left( 2\pi t_{1}^{A} \right) \cos \left(2\pi t_{1}^{A'}\right)     -4 \sum_{\substack{A,B=1\\A>B}}^{8}\cos \left( 2\pi t_{1}^{A} \right) \cos \left( 2\pi t_{1}^{B}\right)  \right. \nonumber\\
&~~~~~~~~~~~~~~~~~~~~~~~~~~~~
\left.  -4 \sum_{\substack{A',B'=9\\A'>B'}}^{16}\cos \left( 2\pi t_{1}^{A'} \right)  \cos \left( 2\pi t_{1}^{B'}\right)   
-24 \right\rbrace .
\end{align}
Note that as we mentioned above, $\Lambda^{(9)}$ is already invariant under the shift $t_{1}^{I}\to t_{1}^{I}+1$ in this model.

Let us study the stability of the Wilson lines from $\Lambda^{(9)}$. The first derivatives of $\Lambda^{(9)}$ are
\begin{align}
\frac{\partial \Lambda^{(9)}}{\partial t^{A}_{1}} &= -64\pi C_{0} \left( \frac{a_0}{\sqrt{\alpha'}} \right)^9 \sin\left(2 \pi t_{1}^{A}\right) \left( \cos\left( 2 \pi t_{1}^{A}\right)-D \right) \\
\frac{\partial \Lambda^{(9)}}{\partial t^{A'}_{1}} &= -64\pi C_{0} \left( \frac{a_0}{\sqrt{\alpha'}} \right)^9 \sin\left( 2 \pi t_{1}^{A'}\right)  \left( \cos\left( 2 \pi t_{1}^{A'}\right)+D \right), 
\end{align}
where we define $D\equiv \sum_{A=1}^{8}\cos\left( 2 \pi t_{1}^{A}\right)-\sum_{A'=9}^{16}\cos\left( 2 \pi t_{1}^{A'}\right)$. The critical points are classified into the following two types:

\begin{enumerate}
	\renewcommand{\labelenumi}{(\roman{enumi})}
	\item For $p+q=p'+q'=8$,
	\begin{equation}\label{first critical}
	t_{1}^{a}=t_{1}^{a'}=0,~~~~t_{1}^{b}=t_{1}^{b'}=\frac{1}{2},
	\end{equation}
	where we denote the indices $a$, $b$, $a'$, $b'$ as
	\begin{equation}
	\begin{split}
	&a=\left\lbrace a_1,\cdots,a_p \right\rbrace,~~b=\left\lbrace b_1,\cdots,b_q \right\rbrace,~~\text{for $1\leq a,b\leq8$}\\
	&a'=\left\lbrace a'_1,\cdots,a'_{p'} \right\rbrace,~~b'=\left\lbrace b'_1,\cdots,b'_{q'} \right\rbrace,
	~~\text{for $9\leq a',b'\leq16$},
	\end{split}
	\end{equation}
	
	\item 
	
	For $0\leq p+q\leq 7, ~0\leq p'+q'\leq 7,~0\leq p+q'\leq 7,~0\leq q+p'\leq 7$,
	\begin{equation}\label{second critical 1}
	t_{1}^{a}=t_{1}^{a'}=0,~~~~t_{1}^{b}=t_{1}^{b'}=\frac{1}{2},
	\end{equation}
	and for $A\neq a,b$, $A'\neq a',b'$,
	\begin{align}\label{second critical 2}
	\cos\left( 2 \pi t_{1}^{A}\right) =-\cos\left( 2 \pi t_{1}^{A'}\right)=D=\frac{p+q'-\left( q+p'\right) }{p+q+p'+q'-15}.
	\end{align}
\end{enumerate}
In order to find out whether the above critical points are stable or not, we need  to evaluate the second derivatives:
\begin{align}
\frac{\partial ^2 \Lambda^{(9)}}{\partial t^{A}_{1} \partial t^{B}_{1}} &= -128\pi^{2} C_{0} \left( \frac{a_0}{\sqrt{\alpha'}} \right)^9\begin{cases}
 \sin\left( 2 \pi t_{1}^{A}\right) 
 \sin\left( 2 \pi t_{1}^{B}\right), &\text{for }A\neq B,\\
 \cos\left( 2 \pi t_{1}^{A}\right) \left( \cos\left( 2\pi t_{1}^A\right)-D \right),&\text{for }A=B,
\end{cases}  \\
\frac{\partial ^2 \Lambda^{(9)}}{\partial t^{A'}_{1} \partial t^{B'}_{1}} &= -128\pi^{2} C_{0} \left( \frac{a_0}{\sqrt{\alpha'}} \right)^9\begin{cases}
 \sin\left( 2 \pi t_{1}^{A'}\right) 
\sin\left( 2 \pi t_{1}^{B'}\right), &\text{for }A'\neq B',\\
 \cos\left( 2 \pi t_{1}^{A'}\right) \left( \cos\left(2 \pi t_{1}^{A'}\right)+D 
 \right),&\text{for }A'=B',
\end{cases}  \\
\frac{\partial ^2 \Lambda^{(9)}}{\partial t^{A}_{1} \partial t^{A'}_{1}} &=128\pi^{2} C_{0} \left( \frac{a_0}{\sqrt{\alpha'}} \right)^9 \sin\left( 2 \pi t_{1}^{A}\right) 
\sin\left( 2 \pi t_{1}^{A'}\right).
\end{align}
 Note that at the critical points (ii),
 \begin{align}
 \frac{\partial ^2 \Lambda^{(9)}}{\partial t^{A}_{1} \partial t^{A}_{1}}= \frac{\partial ^2 \Lambda^{(9)}}{\partial t^{A'}_{1} \partial t^{A'}_{1}}=0,~~~~ \frac{\partial ^2 \Lambda^{(9)}}{\partial t^{A}_{1} \partial t^{A'}_{1}}\neq 0,
 \end{align}
 for $A\neq a,b$, $A'\neq a',b'$. Thus, at least one of the eigenvalues of the Hessian is negative and these critical points are unstable.
   At the critical points (i), the Hessian is a diagonal matrix and the diagonal components are
\begin{align}
-128\pi^{2} C_{0} \left( \frac{a_0}{\sqrt{\alpha'}} \right)^9
 \begin{cases}
1-\left(p+q'-q-p' \right)  & \text{for }~t_{1}^{a},~ t_{1}^{b'},\\
1+\left(p+q'-q-p' \right)  & \text{for }~t_{1}^{b},~ t_{1}^{a'}.
\end{cases}
\end{align}
Therefore, the critical points with $p=q'=8$ or $q=p'=8$, where the gauge symmetry is maximally enhanced to $SO(32)$, are stable. Note that these points are the global minima of $\Lambda^{(9)}$ and there are no local minima.

The cosmological constant is negative at the minima since its leading behavior in the region $a\approx 0$ is written as Eq. \eqref{cc}. There are, however, critical points where $\Lambda^{(9)}\simeq0$, though they are unstable. There are two types of such saddle points. One of them satisfies $\left(P,Q \right)=\left(9,7 \right)$ or $\left(7,9 \right)$ and belongs to the critical points (i). The gauge symmetry is $SO(18)\times SO(14)$ at the points. The other satisfies $\left(P,Q \right)=\left(6,6 \right)$ and belongs to the critical points (ii). The gauge symmetry is $SO(12)\times SO(12)\times (SU(4)\times U(1))$ at the points.

\subsection{Interpolation between $E_8 \times E_8$ and $SO(16)\times SO(16)$}

Let us consider Model I\hspace{-.1em}I with sixteen Wilson lines. According to \eqref{changed point2}, the partition function of Model I\hspace{-.1em}I is 
\begin{align}\label{modelB_16WL}
Z^{(9)}(a,A^I)
&= Z^{(7)}_{B}\left\lbrace \bar{V}_{8}\left(\chi_{OO}^{(0,0)}+\chi_{SS}^{(0,0)}\right) -\bar{S}_{8}\left( \chi_{OS}^{(0,0)}+\chi_{SO}^{(0,0)}\right)\right. \nonumber\\
&~~~~~\left. +\bar{V}_{8}\left( \chi_{OS}^{(1/2,0)}+\chi_{SO}^{(1/2,0)}\right) -\bar{S}_{8}\left( \chi_{OO}^{(1/2,0)}+\chi_{SS}^{(1/2,0)}\right)\right.\nonumber\\
&~~~~~\left. +\bar{O}_{8}\left( \chi_{VC}^{(0,1/2)}+ \chi_{CV}^{(0,1/2)}\right) -\bar{C}_{8}\left( \chi_{VV}^{(0,1/2)}+\chi_{CC}^{(0,1/2)}\right)\right.\nonumber\\
&~~~~~\left. +\bar{O}_{8}\left( \chi_{VV}^{(1/2,1/2)}+\chi_{CC}^{(1/2,1/2)}\right) -\bar{C}_{8}\left( \chi_{VC}^{(1/2,1/2)}+\chi_{CV}^{(1/2,1/2)}\right)\right\rbrace.
\end{align}

\subsubsection{Massless spectrum}

For generic points in the moduli space, the massless spectrum is the same as in Model I: the nine-dimensional gravity multiplet and the gauge bosons of $U(1)^{18}$. Let us search for special points in the moduli space where the additional massless states appear. For simplicity, let us consider the following very special point:
\begin{align}\label{condition2-1}
t_{1}^{A}=\left( \left( 0\right) ^{p_{1}}, \left( 1\right) ^{p_{2}}, \left(\frac{1}{2} \right)^{q_{1}}, \left( -\frac{1}{2} \right)^{q_{2}}  \right),~~~~ t_{1}^{A'}=\left( \left( 0\right) ^{p'_{1}}, \left( 1\right) ^{p'_{2}}, \left(\frac{1}{2} \right)^{q'_{1}}, \left( -\frac{1}{2} \right)^{q'_{2}}  \right),
\end{align}
for $p_{1}+p_{2}+q_{1}+q_{2}=p'_{1}+p'_{2}+q'_{1}+q'_{2}=8$. At this point, we find, from $\bar{V}_{8}\chi_{OO}^{(0,0)}$, the massless vectors with
\begin{empheq}[left={m^{I}=\empheqlbrace}]{alignat=2}
	\label{so(2p)}
	\left( \underline{\pm 1, \pm 1, (0)^{p-2}},(0)^{q};(0)^{8} \right),~~\left( (0)^{p}, \underline{\pm 1, \pm 1, (0)^{q-2}}; (0)^{8}\right) , \\
	\label{so(2p')}
	\left( (0)^{8};  \underline{\pm 1, \pm 1, (0)^{p'-2}},(0)^{q'}\right),~~\left( (0)^{8};  (0)^{p'},\underline{\pm 1, \pm 1, (0)^{q'-2}}\right) ,
\end{empheq}
which correspod to the nonzero roots of $SO(2p)\times SO(2q)\times SO(2p')\times SO(2q')$, where $p= p_1+p_2$, $q= q_1+q_2$, $p'= p'_{1}+p'_{2}$, $q'= q'_{1}+q'_{2}$. From $\bar{S}_{8}\chi_{OO}^{(1/2,0)}$, we find the massless spinors with
\begin{empheq}[left={m^{I}=\empheqlbrace}]{alignat=2}
	\label{bifund_so(2p)}
	\left( \underline{\pm 1,  (0)^{p-1}},\underline{\pm 1,  (0)^{q-1}};(0)^{8} \right)  , \\
	\label{bifund_so(2p')}
	\left((0)^{8}; \underline{\pm 1,  (0)^{p'-1}},\underline{\pm 1,  (0)^{q'-1}} \right) ,
\end{empheq}
which correspond to the $\left( \boldsymbol{2p}, \boldsymbol{2q}\right) $ of $SO(2p)\times SO(2q)$ and the $\left( \boldsymbol{2p'}, \boldsymbol{2q'}\right) $ of $SO(2p')\times SO(2q')$. Furthermore, if $q$ ($q'$) is even, we find the massless states from $\bar{V}_{8}\chi_{SO}^{(1/2,0)}$ and $\bar{S}_{8}\chi_{SO}^{(0,0)}$ ($\bar{V}_{8}\chi_{OS}^{(1/2,0)}$ and $\bar{S}_{8}\chi_{OS}^{(0,0)}$). Let us see what massless states appear when $q=0,2,4,6,8$;
 
 \begin{itemize}
 	\item $q=0$:
 	
 	There are massless  states, in addition to with \eqref{so(2p)}, with
 	\begin{equation}\label{spinor128}
 	m^{A}=\frac{1}{2}\left( \underline{\pm 1,\pm 1,\pm 1,\pm 1,\pm 1,\pm 1,\pm 1,\pm 1}_{+}\right),
 	\end{equation}
 	where the index $+$ ($-$) attached to the underline indicates that the number of pluses is even (odd). If $p_{2}+q_{2}$ (which is just $p_{2}$ in this case) is even, these states come from $\bar{S}_{8}\chi_{SO}^{(0,0)}$. Then, the massless spinors transforming in the $\boldsymbol{128}$ of $SO(16)$ are massless. On the other hand, if $p_{2}+q_{2}$ is odd, the massless states with \eqref{spinor128} come from $\bar{V}_{8}\chi_{SO}^{(1/2,0)}$. Then, the massless vectors with \eqref{so(2p)} and \eqref{spinor128} give the nonzero roots of $E_8$ and there are no massless fermions.
    
 	\item $q=2$:
 	
 	If $p_{2}+q_{2}$ is even, $\bar{V}_{8}\chi_{SO}^{(1/2,0)}$ has the massless vectors with
 	\begin{align}\label{E7su(2)+}
 	m^{A}=\frac{1}{2}\left(\underline{\pm 1,\pm 1,\pm 1,\pm 1,\pm 1,\pm 1}_{+},  \underline{\pm 1, \pm 1}_{+} \right),
 	\end{align}
 	and $\bar{S}_{8}\chi_{SO}^{(0,0)}$ has the massless spinors with
 	\begin{align}\label{E7su(2)-}
 	m^{A}=\frac{1}{2}\left( \underline{\pm 1,\pm 1,\pm 1,\pm 1,\pm 1,\pm 1}_{-}, \underline{\pm 1, \pm 1}_{-} ,\right).
 	\end{align}
 	Noting \eqref{so(2p)} and \eqref{bifund_so(2p)} with $q=2$ correspond to the nonzero roots and the bi-fundamental representation of $SO(12)\times SO(4)$ respectively, we find the gauge bosons of $E_7 \times SU(2)$ and the massless spinors transforming in the $\left( \boldsymbol{56},\boldsymbol{2} \right) $ of $E_7 \times SU(2)$. If $p_{2}+q_{2}$ is odd, then the massless states with \eqref{E7su(2)+} and \eqref{E7su(2)-} come from $\bar{S}_{8}\chi_{SO}^{(0,0)}$ and $\bar{V}_{8}\chi_{SO}^{(1/2,0)}$ respectively, and the massless spectrum is the same as in the $p_2+q_2 \in 2\boldsymbol{Z}$ case.
 	
 	\item $q=4$:
 	
 	If $p_{2}+q_{2}$ is even, $\bar{V}_{8}\chi_{SO}^{(1/2,0)}$ has the massless vectors with
 	\begin{align}\label{SO(8)-}
 	m^{A}=\frac{1}{2}\left( \underline{\pm 1, \pm 1, \pm 1, \pm 1}_{-} ,\underline{\pm 1,\pm 1,\pm 1,\pm 1}_{-}\right),
 	\end{align}
 	and $\bar{S}_{8}\chi_{SO}^{(0,0)}$ has the massless spinors with
 	\begin{align}\label{SO(8)+}
 	m^{A}=\frac{1}{2}\left( \underline{\pm 1, \pm 1, \pm 1, \pm 1}_{+} ,\underline{\pm 1,\pm 1,\pm 1,\pm 1}_{+}\right).
 	\end{align}
 	Using triality symmetry in $SO(8)$, we find the gauge bosons of $SO(16)$ from \eqref{so(2p)} and \eqref{SO(8)-}, and the massless spinors transforming in the $\boldsymbol{128}$ of $SO(16)$ from \eqref{bifund_so(2p)} and \eqref{SO(8)+}. If $p_{2}+q_{2}$ is odd, then the massless states with \eqref{SO(8)+} and \eqref{SO(8)-} come from $\bar{S}_{8}\chi_{SO}^{(0,0)}$ and $\bar{V}_{8}\chi_{SO}^{(1/2,0)}$ respectively, and the massless spectrum is the same as in the $p_2+q_2 \in 2\boldsymbol{Z}$ case.
 
 \item $q=6$:
 
 The massless spectrum is the same as in the $q=2$ case.
 
  \item $q=8$:
 
 The massless spectrum is the same as in the $q=0$ case.
 \end{itemize}
 Table 3 summarizes the massless spectra at the above special points in the moduli space.
 Note that $\bar{V}_{8}\chi_{SO}^{(1/2,0)}$ and $\bar{S}_{8}\chi_{SO}^{(0,0)}$ give the massless states only when $m^{A'}=0$. 
 We can find the massless states with $m^{A}=0$ from $\bar{V}_{8}\chi_{OS}^{(1/2,0)}$, $\bar{S}_{8}\chi_{OS}^{(0,0)}$ and the same massless spectra as in Table 3.
\begin{table}[t] 
	\centering
	\begin{tabular}{|c||c|c|c|c|c|c|} \hline
		Special points &\multicolumn{2}{c|}{$q=0,8$} &\multicolumn{2}{c|}{$q=2,6$} &\multicolumn{2}{c|}{$q=4$} \\ \hline
		$p_2+q_2$ &even&odd&even&odd&even&odd\\ \hline
		Gauge symmetry & $SO(16)$ & $E_8$ & $E_7\times SU(2)$ & $E_7\times SU(2)$ & $SO(16)$ & $SO(16)$\\ \hline 
		Massless spinors & $\boldsymbol{128}$ & $nothing$ & $(\boldsymbol{56},\boldsymbol{2})$ & $(\boldsymbol{56},\boldsymbol{2})$ & $\boldsymbol{128}$& $\boldsymbol{128}$\\ \hline 
	\end{tabular}
	\caption{The special points in the moduli space of Model I\hspace{-.1em}I and the massless spectra at these points.}
\end{table}

In this model, we have restricted our attention to the quite special points \eqref{condition2-1} in the moduli space. We can, of course, find the other special points and identify the massless spectra at the points, by using the expansions given in Appendix \ref{appendix_momentum lattice}.

\subsubsection{Cosmological constant and stability}  
The cosmological constant of Model I\hspace{-.1em}I can be evaluated as 
\begin{align}
\label{modelB_cc}
\Lambda^{(9)}(a,A^I) &\simeq C_{0} \left( \frac{a_0}{\sqrt{\alpha'}} \right)^9   8 \left\lbrace 
128\left( \prod_{A=1}^{8} \cos  \left( \pi t_{1}^{A}\right) + \prod_{A=1}^{8} \sin  \left( \pi t_{1}^{A}\right) \right) \right. \nonumber\\
&~~~~~~~~~~~~~~~~~~~~~~~
\left. +128\left( \prod_{A'=9}^{16} \cos \left( \pi t_{1}^{A'}\right) + \prod_{A'=9}^{16} \sin  \left( \pi t_{1}^{A'}\right) \right) \right. \nonumber\\
&~~~~~~~~~~~~~~~~~~~~~~~
\left. -4 \sum_{\substack{A,B=1\\A>B}}^{8}\cos \left( 2\pi t_{1}^{A} \right) \cos \left( 2\pi t_{1}^{B} \right)  \right. \nonumber\\
&~~~~~~~~~~~~~~~~~~~~~~~
\left.  -4 \sum_{\substack{A',B'=9\\A'>B'}}^{16}\cos \left( 2\pi t_{1}^{A'} \right)  \cos \left( 2\pi t_{1}^{A'} \right)   
-24 \right\rbrace .
\end{align}
Let us analyze the stability of the Wilson lines from $\Lambda^{(9)}$. The first derivatives are
\begin{align}\label{1st derivative model II}
\frac{\partial \Lambda^{(9)}}{\partial t^{A}_{1}} = 64\pi C_{0} \left( \frac{a_0}{\sqrt{\alpha'}} \right)^9\left\lbrace -16 P_{1}^{A} +
 \sin\left(2 \pi t_{1}^{A}\right)\left( D_{1} - \cos\left( 2 \pi t_{1}^{A}\right) \right)\right\rbrace 
\end{align}
where we define
\begin{align}
D_{1}&\equiv \sum_{A=1}^{8}  \cos\left( 2 \pi t_{1}^{A}\right) ,\\
	P_{1}^{A}&\equiv \sin \left( \pi t_{1}^{A} \right) \prod_{\substack{B=1\\B\neq A}}^{8} \cos \left( \pi t_{1}^{B}\right) -\cos\left( \pi t_{1}^{A}\right)  \prod_{\substack{B=1\\B\neq A}}^{8} \sin \left( \pi t_{1}^{B} \right) .
\end{align}
Note that the first derivatives \eqref{1st derivative model II} do not depend on $t_{1}^{A'}$ and
$\partial\Lambda^{(9)}/\partial t_{1}^{A'}$ takes the same form as in \eqref{1st derivative model II}. So, it is sufficient to analyze the stability  with $t_{1}^{A}$ only. Since it is difficult to solve $\partial\Lambda^{(9)}/\partial t_{1}^{A}=0$ for generic values of $t_{1}^{A}$, we use the ansatz that $t_{1}^{A}$ at critical points can be written as the following form, except the freedoms of permutations of the components:
\begin{align}
t_{1}^{A}=\left( (0)^{p_{1}}, (1)^{p_{2}}, \left( \frac{1}{2}\right) ^{q_{1}}, \left( \frac{1}{2}\right) ^{q_{2}}, \left( \frac{1}{4}\right) ^{r_{1}}, \left( -\frac{3}{4}\right) ^{r_{2}},  \left( -\frac{1}{4}\right) ^{s_{1}},  \left( \frac{3}{4}\right) ^{s_{2}} \right) ,
\end{align}
where $p+q+r+s=8$ for $p=p_{1}+p_2$, $q=q_{1}+q_2$, $r=r_{1}+r_2$, $s=s_{1}+s_2$. Then, we find that the critical points have to satisfy at least one of the following conditions:
\begin{enumerate}
	\renewcommand{\labelenumi}{(\roman{enumi})}
	\item $p\geq 2,~q\geq 2$ and $p+q=8$; 
	\item $p\geq 2,~q\geq 2$ and $p=q$;
	\item $p=q=0$ and $s$ is even;
	\item $\left(p,q \right) = (0,2),(0,4)$ and $q_{2}+r_{2}+s_{1}$ is even;
	\item $\left(p,q \right) = (2,0),(4,0)$ and $p_{2}+r_{2}+s_{2}$ is even;
	\item $\left(p,q \right) = (8,0),(0,8)$.
\end{enumerate}
The second derivatives are 
\begin{align}
\frac{\partial ^2 \Lambda^{(9)}}{\partial t^{A}_{1} \partial t^{B}_{1}} &= 128\pi^{2} C_{0} \left( \frac{a_0}{\sqrt{\alpha'}} \right)^9\begin{cases}
 \left\lbrace -8 P_{0}+  \cos \left( 2\pi t_{1}^{A} \right) \left( D_{1}
- \cos \left( 2\pi t_{1}^{A} \right)\right) \right\rbrace , &\text{for }A= B,\\
 \left\lbrace -8 P_{2}^{AB}+  \sin \left( 2\pi 
t_{1}^{A} \right)  \sin \left( 2\pi 
t_{1}^{B}\right)  \right\rbrace,&\text{for }A\neq B,
\end{cases}  
\end{align}
where
\begin{align}
P_{0} &\equiv \prod_{A=1}^{8} \cos \left( \pi t_{1}^{A} \right)  + \prod_{A=1}^{8} \sin \left( \pi t_{1}^{A} \right) ,\\
P_{2}^{AB} &\equiv \sin \left( \pi t_{1}^{A} \right)\sin \left( \pi t_{1}^{B}\right) \prod_{\substack{C=1\\C \neq A,B}}^{8} \cos \left( \pi t_{1}^{C} \right) + \cos \left( \pi t_{1}^{A} 
\right) \cos \left( \pi t_{1}^{B}\right) \prod_{\substack{C=1\\C \neq A,B}}^{8} \sin \left( \pi t_{1}^{C} \right).
\end{align}
Let us evaluate the Hessian at the critical points and analyze the stability of the Wilson lines.
Note that if a symmetric matrix is positive definite, then all of the leading principal minors must be positive. Then, it turns out that the critical points which satisfy one of the five conditions (i)-(v) are unstable. At the critical points satisfying the condition (vi), the second derivatives are
\begin{align}
\frac{\partial ^2 \Lambda^{(9)}}{\partial t^{A}_{1} \partial t^{A}_{1}} &=128\pi^{2} C_{0} \left( \frac{a_0}{\sqrt{\alpha'}} \right)^9 \begin{cases}
\left( 8\left( -1\right) ^{p_{2}+1} +7 \right)  , &\text{for }\left( p,q\right)=(8,0) ,\\
 \left( 8\left( -1\right) ^{q_{2}+1} +7 \right) ,&\text{for }\left( p,q\right)=(0,8) ,
\end{cases}  \\
\frac{\partial ^2 \Lambda^{(9)}}{\partial t^{A}_{1} \partial t^{B}_{1}} &=0, ~~\text{for }A\neq B.
\end{align}
Therefore, the critical points with $\left( p,q\right)=(8,0)$, $p_{1}\in2\boldsymbol{Z}+1$ or $\left( p,q\right)=(0,8)$, $q_{1}\in2\boldsymbol{Z}+1$, where the gauge symmetry is maximally enhanced to $E_{8}$, have the positive definite Hessian. Taking into account the derivatives with respect to $t_{1}^{A'}$, we find that the minima of $\Lambda^{(9)}$ correspond to the points where the gauge symmetry is $E_{8}\times E_{8}$. There are no massless fermions.

At the minima of $\Lambda^{(9)}$, the cosmological constant is negative since its leading term is proportional to $n_{F}-n_{B}$.
As in Model I, however, we can find two types of critical points with $\Lambda^{(9)}\simeq 0$, which are unstable. One of them is a set of critical points where the gauge symmetry is $SO(16)\times SO(10)\times SO(6)$. At the other type of the critical points, the gauge symmetry is $\left( SU(8)\times U(1) \right)^2 \times \left( SU(2)\times U(1) \right)^2$.

\section{Conclusions and Discussions}\label{conclusion}
We have calculated the partition function of 9D interpolating models under the Wilson line backgrounds and studied the massless spectra. In this work, we consider two interpolations between 10D heterotic string models:
\begin{itemize}
	\item Model I: 10D SUSY $SO(32)$ model $\longleftrightarrow$ 10D $SO(16)\times SO(16)$ model
	\item Model I\hspace{-.1em}I: 10D SUSY $E_8 \times E_8$ model $\longleftrightarrow$ 10D $SO(16)\times SO(16)$ model
\end{itemize}
Although the gauge symmetry is $U(1)^{18}$ and all of the fermions are massive at generic points in the moduli space, we have found some special points where the gauge symmetry is enhanced to non-Abelian groups and some fermions are massless. We have evaluated the cosmological constants in the region $a\approx 0$ as functions of moduli and analyze the stability of the Wilson lines. It turns out that the Wilson lines are stabilized when the gauge symmetry is maximally enhanced, and the cosmological constants are negative at the stable points.

In this paper, we have focused only on 9D interpolating models constructed from 10D endpoint models with the Scherk-Schwarz compactification. In order to construct more realistic models, of course, we have to compactify more dimensions and consider 4D string models. It is not difficult to obtain 4D string models by compactifying the 9D interpolating models studied in this paper on tori. In such 4D models, the process of supersymmetry breaking by the Scherk-Schwarz mechanism is $\mathcal{N} =4\to 0$. Rather, it is more interesting to consider the supersymmetry breaking $\mathcal{N} =2\to 0$ or $\mathcal{N} =1\to 0$ by compactifying on orbifolds as in Ref. \cite{Kounnas:2015yrc,Kounnas:2016gmz,Kounnas:2017mad}, in order to investigate the phenomenological aspects of interpolating models. 

In Ref. \cite{Blum:1997gw,Blum:1997cs}, S-duality between heterotic string vacua and type I string vacua in non-supersymmetric cases is explored. In Ref. \cite{Abel:2018zyt,Partouche:2019pgv,Angelantonj:2019gvi,Abel:2020ldo}, the moduli space in non-supersymmetric type I models constructed by the Scherk-Schwartz compactification, which is characterized by brane configurations, are discussed. On the other hand, the moduli in heterotic picture correspond to the parameters of boosts of the momentum lattices. It is interesting to figure out the correspondence between the moduli space and the landscape in the different pictures explicitly, as well in \cite{Dienes:2012dc}.

In this work, we focused on the interpolations in which the non-supersymmetric endpoint is the tachyon-free heterotic $SO(16)\times SO(16)$ model. There are a lot of 10D non-supersymmetric string models that have a tachyonic state, and such tachyonic string models can be good starting points in non-supersymmetric string phenomenology\cite{Faraggi:2019fap,Faraggi:2019drl}.
It is interesting to consider the interpolations from tachyonic string vacua to supersymmetric ones because the tachyon becomes massive as the radius approaches the region where supersymmetry is asymptotically restoring.

Important future work is concerned with the stability of the moduli. The cosmological constants are exponentially suppressed at the saddle points while they are negative at the minima of the one-loop effective potentials. The question of whether de Sitter string vacua can be constructed or not is one of the recent main issues in string phenomenology. It has been conjectured that string vacua with positive cosmological constants are not allowed\cite{Vafa:2005ui,Ooguri:2006in,Obied:2018sgi,Ooguri:2018wrx,Garg:2018reu,Ferrara:2019tmu,Banerjee:2020wix}. In spite of this conjecture, our universe has a very small positive cosmological constant. So, constructions of metastable string vacua have been attracted attention recently.
One of the possibilities to obtain such metastable vacua is to include effects of higher loop corrections. As in the KKLT scenario\cite{Kachru:2003aw}, the minima of the effective potential might be uplifted due to the higher loop corrections and de Sitter or Minkowski metastable vacua might be realized.
\section*{Acknowledgments}

We thank  J. X. Lu, Shun'ya Mizoguchi and Takao Suyama for helpful discussion on this subject.
We thank the organizers of Conference on Recent Developments in Strings and Gravity of the Corfu Summer Institute 2019 for the opportunity to present a series of our work. The work of H. I. was partially supported by JSPS KAKENHI Grant Number 19K03828.


\appendix

\section{Notation}\label{notations}

We summarize the notation for some functions that appear in the partition functions. The Dedekind eta function is defined as
\begin{equation}
\eta(\tau)=q^{1/24}\prod_{n=1}^{\infty}\left( 1-q^{n}\right),
\end{equation} 
where $q=e^{2\pi i\tau}$. The theta function with characteristics is defined as
\begin{equation}
\vartheta
\begin{bmatrix} 
\alpha\\ 
\beta\\ 
\end{bmatrix}(z,\tau)=\sum_{n=-\infty}^{\infty}\exp\left( \pi i (n+\alpha)^2 \tau +2\pi i (n+\alpha)(z+\beta) \right). 
\end{equation}
The $SO(2n)$ characters are defined in terms of the Dedekind eta function and the theta functions as follows:
\begin{align}
\left(
\begin{array}{c}
O_{2n}\\
V_{2n}
\end{array}
\right)&\equiv\frac{1}{2\eta^{n}}\left( \vartheta^{n}
\begin{bmatrix} 
0\\ 
0\\ 
\end{bmatrix}(0,\tau)\pm \vartheta^{n}
\begin{bmatrix} 
0\\ 
1/2\\ 
\end{bmatrix}(0,\tau)
\right)= \frac{1}{\eta^{n}} \sum_{m^{A}\in \boldsymbol{Z}_{\pm}^{n}}q^{|m|^2/2},\\
\label{SC}\left(
\begin{array}{c}
S_{2n}\\
C_{2n}
\end{array}
\right)&\equiv\frac{1}{2\eta^{n}}\left( \vartheta^{n}
\begin{bmatrix} 
1/2\\ 
0\\ 
\end{bmatrix}(0,\tau)\pm \vartheta^{n}
\begin{bmatrix} 
1/2\\ 
1/2\\ 
\end{bmatrix}(0,\tau)
 \right)=\frac{1}{\eta^{n}} \sum_{m^{A}\in \boldsymbol{Z}_{\pm}^{n}+1/2}q^{|m|^2/2},
\end{align}
where $|m|^2=\sum_{A=1}^{n}\left( m^{A}\right) ^{2}$, and $\boldsymbol{Z}_{\pm}^{n}$ is defined as
\begin{align}
\boldsymbol{Z_+}^8=\left\lbrace m^A \in \boldsymbol{Z}^n~\left| ~\sum_{A}m^A\in  2\boldsymbol{Z}\right. \right\rbrace ,~~~~\boldsymbol{Z_-}^n=\left\lbrace m^A \in \boldsymbol{Z}^n~\left| ~\sum_{A}m^A\in  2\boldsymbol{Z}+1\right. \right\rbrace.
\end{align}
Note that in the second equality of \eqref{SC}, we assume $n\in 4\boldsymbol{Z}$.

\section{Examples of 10D endpoint models}\label{endpoints}

In this appendix, we provide concrete examples of $Z_{+}^{+}$ in Eq. \eqref{Z_int} and the $\boldsymbol{Z}_2$ action $Q$ which relates $Z_{+}^{+}$ to $\left( Z_{+}^{+} +Z_{-}^{+}+Z_{+}^{-}+Z_{-}^{-}\right)/2$. This review is based on Ref. \cite{Blum:1997gw}. In Model I which interpolates from the supersymmetric $SO(32)$ model to the $SO(16)\times SO(16)$ model, $Z_{+}^{+}$ is given by
\begin{align}
Z_{+}^{+}=\left( \bar{V}_{8}-\bar{S}_{8} \right)\left(O_{16} O_{16}+V_{16}V_{16}+S_{16}S_{16}+C_{16}C_{16}\right),
\end{align}
which is zero due to the Jacobi's abstruse identity. In order to obtain the $SO(16)\times SO(16)$ model as the other endpoint, we adopt a $\boldsymbol{Z}_{2}$ action $Q$ as $\bar{R}_{SC}R_{SC}^{(1)}R_{VS}^{(2)}$, where $\bar{R}_{SC}$ acts the right-moving $SO(8)$ representations and  $R_{SC}^{(1)}$ and $R_{SV}^{(2)}$ act the first and second left-moving $SO(16)$ representations respectively as follows:
\begin{align}
\bar{R}_{SC}&:\left( \bar{O}_{8}, \bar{V}_{8}, \bar{S}_{8}, \bar{C}_{8}\right) \to \left( \bar{O}_{8}, \bar{V}_{8}, -\bar{S}_{8}, -\bar{C}_{8}\right), \\
R_{SC}^{(1)}&:\left( O_{16}, V_{16}, S_{16}, C_{16}\right) \to \left( O_{16}, V_{16}, -S_{16}, -C_{16}\right), \\
R_{SV}^{(2)}&:\left( O_{16}, V_{16}, S_{16}, C_{16}\right) \to \left( O_{16}, -V_{16}, -S_{16}, C_{16}\right).
\end{align}
By projecting out $Z_{+}^{+}$ by $Q$ and using the transformations of $SO(2n)$ characters under $S:\tau \to -1/\tau$
\begin{equation}
\left(
\begin{array}{c}
O_{2n}\\
V_{2n}\\
S_{2n}\\
C_{2n}\\
\end{array}
\right)
\to  \left(
\begin{array}{cccc}
1&1&1&1\\
1&1&-1&-1\\
1&-1&i^n&-i^n\\
1&-1&-i^n&i^n\\
\end{array}
\right)
\left(
\begin{array}{c}
O_{2n}\\
V_{2n}\\
S_{2n}\\
C_{2n}\\
\end{array}
\right),
\end{equation}
we obtain 
\begin{align}\label{so16so16}
\frac{1}{2}\left( Z_{+}^{+} +Z_{-}^{+}+Z_{+}^{-}+Z_{-}^{-}\right) =& \bar{O}_{8} \left(V_{16}C_{16}+C_{16}V_{16} \right)+\bar{V}_{8} \left(O_{16}O_{16}+S_{16}S_{16} \right)\nonumber\\ 
&-\bar{S}_{8} \left(V_{16}V_{16}+C_{16}C_{16} \right)-\bar{C}_{8} \left(O_{16}S_{16}+S_{16}O_{16} \right), 
\end{align}
which is the partition function of the $SO(16)\times SO(16)$ model except for the factor $Z_{B}^{(8)}$.

In Model I\hspace{-.1em}I, in which the supersymmetric endpoint is the $E_{8}\times E_{8}$ model, $Z_{+}^{+}$ is given by
\begin{align}
Z_{+}^{+}=\left( \bar{V}_{8}-\bar{S}_{8} \right)\left(O_{16} +S_{16}\right)\left(O_{16} +S_{16}\right).
\end{align}
Taking the same $\boldsymbol{Z}_2$ action $Q$ as in Model I, we can obtain Eq. \eqref{so16so16} from $Z_{+}^{+}$ by projecting out by $Q$ and adding the twisted sectors.

In order to construct 9D interpolating models, $Z_{B}^{(8)}Z_{+}^{+}$ have to be compactified on a circle with the twist, in addition to by $Q$, by a half translation $\mathcal{T}$:
 \begin{align}
 \mathcal{T}:~X^{9}\to X^{9} + \pi R.
 \end{align}
As the states which are invariant under $\mathcal{T}$ have even numbers of quantized internal momenta, the effects of the projection by $\mathcal{T}$ appear in the momentum lattices. The compactification of $Z_{+}^{+}$ on a circle with the total $\boldsymbol{Z}_2$ twist by $\mathcal{T}Q$ provides the partition function which takes the form \eqref{Z_int}.

\section{Suppression of the cosmological constant in the asymptotic region $a\approx 0$}\label{calculation of cc}
In this appendix, we review, in the current notation, the basic argument and derivation of the suppression of the cosmological constant \cite{Itoyama:1986ei,Itoyama:1987rc} for a generic interpolating model in the region $a\approx 0$ where supersymmetry is asymptotically restoring. For definiteness, we demonstrate this for the 9D interpolating models discussed in the text of this paper, but the derivation  is applicable to more general $D$-dimensional interpolating models.

As is well known, the cosmological constant in $D$-dimensions is written as the integral of the partition function
\begin{align}
\Lambda^{(D)} = -\frac{1}{2} \left( 4\pi^2 \alpha'\right)^{-D/2} \int_{\mathcal{F}}  \frac{d^2 \tau}{\tau_{2}^2} Z^{(D)}
\end{align}
over the fundamental region $\mathcal{F}$ of the modular group
\begin{align}
\mathcal{F}=\left\lbrace\tau=  \tau_{1}+ i\tau_{2}\in \boldsymbol{C}~ \left|-\frac{1}{2}\leq \tau_{1}\leq \frac{1}{2},~|\tau|\geq 1 \right. \right\rbrace.
\end{align}
For our convenience, we decompose $\mathcal{F}$ into two pieces $\mathcal{F}_{\geq 1}=\mathcal{F}|_{\tau_{2}\geq 1}$ and $\mathcal{F}_{< 1}=\mathcal{F}|_{\tau_{2}< 1}$.
Let us take Eq. \eqref{Z_int} as $Z^{(D)}$ and look at the region. By the assumption made in this paper, $Z_{+}^{+}$ is zero and the states with non-vanishing winding numbers are exponentially suppressed. This permits us to write
\begin{align}\label{eq84}
Z^{(9)}_{\text{int}}&\simeq \frac{1}{2}Z_{B}^{(7)}Z^{+}_{-}\left( \Lambda_{0,0}- \Lambda_{1/2,0}\right)\nonumber\\ 
&=\frac{\sqrt{\tau_{2}}}{2}Z^{(8)}_{B}Z^{+}_{-}\sum_{n\in \boldsymbol{Z}}\left( e^{-4\pi \tau_{2} a^{2} n^{2}}-e^{-4\pi \tau_{2} a^{2} \left(n+1/2 \right) ^{2}} \right).
\end{align}
The physical meaning of this expression involving the momentum sum with an alternating sign is rather clear; the mass splitting $\alpha' M_{s}^{2}=a^{2}$ between a boson and a fermion adjacent to each other is small and a series of nearly degenerate bose-fermi supermultiplets are formed. While this expression itself does not allow us to estimate its value in the region that we work with, we can, of course, invoke the Jacobi imaginary transformation to recast Eq. \eqref{eq84} into 
\begin{align}
Z^{(9)}_{\text{int}}\simeq \frac{1}{a} Z^{(8)}_{B}Z^{+}_{-}\sum_{n=1}^{\infty}e^{-(2n-1)^{2}\pi/\left( 4 \tau_{2} a^2\right)   }.
\end{align}
Let us show the exponential suppression on the contribution to the cosmological constant from a generic non-vanishing level $m\neq 0$, taking a factor $q^{m_+}\bar{q}^{m_-}\equiv \left(q\bar{q} \right)^{m}e^{2\pi i \tau_{1} s} $ from $Z_{B}^{(8)}Z^{+}_{-}$. Using the inequality on the arithmetic-geometric mean, we obtain 
\begin{align}
\sum_{n=1}^{\infty}e^{-(2n-1)^{2}\pi/\left( 4 \tau_{2} a^2\right)   }\left( q\bar{q} \right)^{m} \leq \frac{e^{-2\pi \sqrt{m}/a}}{1-e^{-4\pi \sqrt{m}/a}}.
\end{align}
This bound is $\tau_{2}$ independent and together with $\tau_{2}^{-6}$, it can be integrated over $\mathcal{F}_{\geq 1}$, giving a finite prefactor. So, this part of the contribution is at least suppressed by an exponential factor $e^{-2\pi/a}$. 
As for the integration over $\mathcal{F}_{<1}$, the domain itself is finite and the integrand itself is singularity free. We can easily bound the integrand $Z_{\text{int}}^{(9)}$ by
\begin{align}
Z_{\text{int}}^{(9)}< \frac{1}{a}Z_{B}^{(8)}Z_{-}^{+}\sum_{n=1}^{\infty} e^{-(2n-1)\pi/\left( 4 \tau_{2} a^2 \right)  }=\frac{1}{a}Z_{B}^{(8)}Z_{-}^{+}\frac{e^{-\pi /\left( 4a^2\right) }}{1-e^{-2\pi /\left( 4a^2\right) }}
\end{align}
So, the contribution from this part of the integration to the cosmological constant is suppressed at least by an exponential factor $e^{-\pi/4a^2}$.

Let us now turn out attention to the $m=0$ case, namely, the contribution from the massless levels. The same reasoning holds for the integration over $\mathcal{F}_{<1}$ as in the $m\neq 0$ case and the contribution from this part of the integration to the cosmological constant is exponentially suppressed as well. Finally, let us see the contribution from the integration over $\mathcal{F}_{\geq 1}$. We need to evaluate the following integral up to an exponential accuracy $e^{-\pi/4a^2}$ at $a \approx 0$:
\begin{align}
\frac{1}{a}\int_{1}^{\infty} d\tau_{2} \tau_{2}^{-6} \sum_{n=1}^{\infty} e^{-(2n-1)^2/\left( 4\tau_{2} a^{2}\right) }\simeq \frac{4!}{\pi^{5}}a^{9}\zeta\left( 10,1/2 \right),
\end{align}
where $\zeta\left( 10,1/2 \right)=\left( 2^{10}-1\right)\zeta(10).$ From these estimates, we conclude that
\begin{align}
\Lambda^{(9)}=\frac{4!}{\pi^{5}}a^{9}\zeta\left( 10,1/2 \right) \left( n_{F}-n_{B} \right) \left(4\pi^2 \alpha' \right)^{-9/2}+\mathcal{O}\left( e^{-2\pi/a} \right) .
\end{align}

In Ref. \cite{Abel:2015oxa}, the subleading contributions to the cosmological constant have been derived.

\section{Momentum lattices boosted by Wilson lines}\label{appendix_momentum lattice}

In order to see the degrees of freedom of the states at each mass level from the partition functions \eqref{modelA_16WL} and \eqref{modelB_16WL}, let us see the behaviors
of the momentum lattice \eqref{boosted lattices} in the region $a\approx 0$:
\begin{align}\label{lattice_zerobeta}
\Lambda^{(\alpha,0)}\begin{bmatrix} 
\gamma_{1}&\gamma_{2}\\ 
\delta_{1}&\delta_{2}\\ 
\end{bmatrix}&\underset{w=0}{\simeq} \left( \bar{\eta} \eta\right) ^{-1} \eta^{-16} \sum_{n\in \boldsymbol{Z}}\sum_{m^A\in \boldsymbol{Z}^8}\sum_{m^{A'}\in \boldsymbol{Z}^8}\left(  \prod_{A=1}^{8} e^{2\pi \delta_{1}\left( m^A + \gamma_{1} \right)} q^{\left(m^A+\gamma_{1} \right)^2/2  } \right) \nonumber\\
&~~~~~~~~~~~~~\times\left( \prod_{A'=9}^{16}e^{2\pi \delta_{2}\left( m^{A'} + \gamma_{2} \right)} q^{\left(m^{A'}+\gamma_{2} \right)^2/2  } \right) \left(q\bar{q} \right) ^{a_{0}^2\left( n+\alpha +t_{1}\cdot(m+\gamma)\right)^2 }
\end{align}
By using this, the following products, which contain the contributions from the left-moving states to the partition function, are expanded as
\begin{align}
\bar{\eta}\eta^{-7} \chi_{**}^{(\alpha,0)}
&\simeq   \eta^{-24}\sum_{n\in \boldsymbol{Z}}\sum_{m^A}\sum_{m^{A'}} 
q^{|m|^2/2}\left( q\bar{q}\right) ^{a_{0}^2\left( n+\alpha +m\cdot t_{1} \right)^2 }.
\end{align}
Here the sums over $m^A$ and $m^{A'}$ depend on $\chi_{**}^{(\alpha,0)}$ as in the following table:
 \begin{table}[H] 
	\centering
	\begin{tabular}{|c|c|c|c|c|c|c|c|c|} \hline
		 &$\chi_{OO}^{(\alpha,0)}$ & $\chi_{VV}^{(\alpha,0)}$ & $\chi_{SS}^{(\alpha,0)}$ &$\chi_{CC}^{(\alpha,0)}$ &$\chi_{SO}^{(\alpha,0)}$ &$\chi_{OS}^{(\alpha,0)}$ &$\chi_{CV}^{(\alpha,0)}$ &$\chi_{VC}^{(\alpha,0)}$ \\ \hline 
		 $m^{A}$&$\boldsymbol{Z_+}^8$ & $\boldsymbol{Z_-}^8$ & $\boldsymbol{Z_+}^8+1/2$ &$\boldsymbol{Z_-}^8+1/2$ &$\boldsymbol{Z_+}^8+1/2$ &$\boldsymbol{Z_+}^8$ &$\boldsymbol{Z_-}^8+1/2$ &$\boldsymbol{Z_-}^8$ \\ \hline 
		 $m^{A'}$&$\boldsymbol{Z_+}^8$ & $\boldsymbol{Z_-}^8$ & $\boldsymbol{Z_+}^8+1/2$ &$\boldsymbol{Z_-}^8+1/2$ &$\boldsymbol{Z_+}^8$ &$\boldsymbol{Z_+}^8+1/2$ &$\boldsymbol{Z_-}^8$ &$\boldsymbol{Z_-}^8+1/2$ \\ \hline 
	\end{tabular}
\end{table}\noindent
Note that 
\begin{align}
&\frac{1}{2}\left( \underline{\pm 1,\pm 1,\pm 1,\pm 1,\pm 1,\pm 1,\pm 1,\pm 1}_{+}\right)\in \boldsymbol{Z_+}^8+\frac{1}{2} ,\\
&\frac{1}{2}\left( \underline{\pm 1,\pm 1,\pm 1,\pm 1,\pm 1,\pm 1,\pm 1,\pm 1}_{-}\right)\in \boldsymbol{Z_-}^8+\frac{1}{2},
\end{align}
which correspond to the $\boldsymbol{128}$ and the $\boldsymbol{128'}$ representations of $SO(16)$ respectively. The prefactor $\eta^{-24}$, which represents the contributions from the nonzero modes of the left-moving bosonic string coordinates, is expanded as
\begin{align}
\eta^{-24}=q^{-1} + 24 + \mathcal{O}(q).
\end{align}
Then, for generic values of $t_{1}^I$, we find the massless states with $n=m^I=0$ from $\bar{\eta}\eta^{-7} \chi_{OO}^{(0,0)}$, whose left-moving degrees of freedom are 24. There are no other massless states for generic values of $t_{1}^I$. However, if $t_{1}^{I}$ satisfies $m\cdot t_{1}\in \boldsymbol{Z} +\alpha $ for $m^{I}$ satisfying $|m|^2=2$, then there are additional massless states. For example, from $\chi_{VV}^{(\alpha,0)}$, we find the massless states with
\begin{align}
m^{I}=\left(m^{A}; m^{A'} \right) =\left( \underline{\pm 1, (0)^{7}}; \underline{\pm 1, (0)^{7}}\right),
\end{align}
if all the components of $t_{1}^{I}$ are integers or half-integers. We can find the other special values of $t_{1}^{I}$ and the additional massless states at the values which are given in Sect. \ref{General 9D interpolating model}.
 Note that the states found from $\chi_{SS}^{(\alpha,0)}$, $\chi_{CC}^{(\alpha,0)}$, $\chi_{VC}^{(\alpha,0)}$, $\chi_{CV}^{(\alpha,0)}$ can not be massless because there is no $m^{I}$ satisfying $|m|^2=2$ in the sums. 

\end{document}